\documentclass[preprint,nofootinbib,aps,superscriptaddress,eqsecnum]{revtex4-1}
 \pdfoutput=1
\usepackage{amsmath,amssymb,graphicx,subfigure}

\usepackage{float}
\usepackage{graphicx}
\usepackage{amsmath}
\usepackage{amsfonts}
\usepackage{amssymb}
\usepackage{hyperref}
\usepackage{subfigure}
\usepackage{slashed}
 \usepackage{setspace} 
\usepackage{caption,ulem}
\usepackage{color}
\allowdisplaybreaks
\def\bea{\begin{eqnarray}}
\def\eea{\end{eqnarray}}
 \def\be{\begin{equation}}
\def\ee{\end{equation}}
\def\nn{\nonumber}
\def\mpl{M_{\rm Pl}}
\newcommand{\beq}{\begin{equation}}
\newcommand{\eeq}{\end{equation}}

\def\nn{\nonumber}
\def\mpl{M_{\rm Pl}}
\usepackage[utf8]{inputenc}

\DeclareUnicodeCharacter{2212}{-}

\def\nn{\nonumber}

\def\ra{\rightarrow}
\def\mpl{M_{\rm Pl}}

\def\mpl{M_{\rm Pl}}

\begin{document}
\title{FIMP DM in the Extended Hyperchargeless Higgs Triplet Model }
\author{Pritam Das}
\email{prtmdas9@gmail.com}
\author{Mrinal Kumar Das}
\email{mkdas@tezu.ernet.in}
\affiliation{Department of Physics, Tezpur University, Assam-784028, India.}
 \author{Najimuddin Khan}\email{psnk2235@iacs.res.in} 
\affiliation{School of Physical Sciences, Indian Association for the Cultivation of Science 2A $\&$ 2B, Raja S.C. Mullick Road, Kolkata 700032, India.}
\affiliation{School of physics, Institute for Research in Fundamental Sciences (IPM), P.O.Box 19395-5531, Tehran, Iran.\vspace{1.80cm}}


\begin{abstract}\vspace*{10pt}
We perform an exclusive study on the Feebly Interacting Massive Particle (FIMP)  dark matter candidate in an extended hyperchargeless ($Y=0$) Higgs triplet model. 
The additional $Z_2$ odd neutral fermion singlet plays the role of dark matter with support from two other vector-like fermion doublets. The mixing between the neutral component of a doublet and singlet fermions controls the current relic density through the Freeze-in mechanism, whereas the additional doublet fermion helps to get the neutrino mass and mixing angles.
We obtain a broad region of the parameter spaces satisfying the current relic density and neutrino mass and mixing angles.

\end{abstract}

\maketitle

\section{Introduction}
Through meticulous measurements of cosmic microwave background (CMB) anisotropies, cosmology based experiments such as  PLANCK \cite{Aghanim:2018eyx}, WMAP \cite{Bennett:2012zja} have
suggested the existence
dark matter (DM), giving rise to around 26\% of the
present universe’s energy density. Astrophysical pieces of evidence like galaxy cluster observations by F. Zwicky \cite{Zwicky:1937zza}, observations of galaxy rotation curves\cite{Freese:2008cz} to the more recent observation of the bullet cluster \cite{Clowe:2006eq} also agrees with the cosmological observations. Now, one can explain the dark matter relic density as, $\Omega h^2=0.1198\pm 0.0026$~\cite{Aghanim:2018eyx} using various theories~\cite{Kolb:1990vq, Hall:2009bx}. Interestingly, all those evidence are based on gravitational interactions only, and until date, not a single evidence has reported in support of the particle interaction of DM. The cosmological and astrophysical data can only explain how much dark matter is there in the universe, but to understand what it is, one must look from the particle physics perspective. The explanation from the particle physics perspective is quite obvious, and no SM particle has the quality to fulfill the criteria to be a DM candidate. 
There is still no general agreement on what it is composed of.
The non-baryonic dark matter component can be put into three categories based on their velocities, e.g., hot dark matter (HDM), warm dark matter (WDM), and cold dark matter (CDM). The neutrino has high velocities as an HDM is ruled out~\cite{Murayama:2007ek}. The WDM and CDM have moderate velocities that have explained the dark matter densities and other experimental constraints.

One of the most intriguing experimental findings in particle physics is the phenomenon of neutrino oscillation.
The atmospheric, solar, reactor and accelerator neutrino oscillation experiments~\cite{Abe:2016nxk, An:2012eh, Abe:2011fz} have predicted that the three flavor of the neutrinos mix among themselves and they have a tiny mass, which is the only solid experimental evidence available in favour of the beyond the standard model (BSM) framework. The oscillation experiments are only sensitive to the mixing angles and mass square differences ($\Delta m_{ij}^2=m_i^2-m_j^2$). 
From various observations, we get a constraints on the sum of the all neutrino mass eigenvalues ($\sum_{m_i}<0.117$ eV \cite{Choudhury:2018byy}, with $i=1,2,3$).

From the SM point of view, there is no sufficient candidate left to propose as a dark matter candidate or explain the neutrino mass. Therefore, one must head towards BSM frameworks to explain the current dark matter density and neutrino low energy variables i.e. mass and mixing angles.
The recent LHC Higgs signal strength data~\cite{Sirunyan:2017khh, Sirunyan:2018koj} also allows us to include additional fields of the new physics beyond the SM. 
{Among the various BSM scenarios that are proposed in the literature to explain the tiny neutrino masses,the seesaw mechanisms are the most popular one} \cite{Mohapatra:1979ia,Bernabeu:1987gr}.
It is known that in order to get a neutrino mass of the sub-eV scale, one has to take the new particles to be extremely heavy (right-handed neutrino) or else take the new couplings to be extremely small or need a vast fine-tuning in the new Yukawa sector.

A model that can address both light neutrino mass and dark matter with a minimum particle content is much appealing and well-motivated in the model building business.
A working model is considered to be completed when it can simultaneously explain light neutrino observables and dark matter, and both sectors are well connected. There are many popular frameworks in the literature \cite{Babu:2009fd,Borah:2018gjk,Khan:2012zw,Das:2019ntw,Bhattacharya:2017fid,Ghosh:2017fmr,Das:2020hpd,Das:2021xat, Das:2014fea,Barry:2011wb,Burgess:2000yq,Cohen:2011ec,Grimus:2009mm,Dev:2013yza,Toma:2013zsa}. 
The framework proposed by E.Ma \cite{Ma:2006km}, that gained severe attention in accommodating both dark matter and neutrino mass at loop level is known as the $scotogenic$ model, where the dimension-5 operator is realized at the one-loop level \cite{weinberg}.
The notable feature of this framework is the way it connects neutrino and DM. Due to the additional $Z_2$ discrete symmetry, new fields contributing to the loop to produce sizable neutrino mass acquire opposite parity to the SM fields. Hence, The new field becomes stable due to odd $Z_2$ charge imposed on it, and can be addressed as a viable dark matter candidate. With these impressive features in addressing neutrino mass and dark matter, the scotogenic model has gained popularity in recent days~\cite{GonzalezFelipe:2003fi,Fraser:2014yha, Merle:2015ica, Law:2013saa,Mahanta:2019gfe, Klein:2019iws, Sarma:2020msa}.

Many of the DM genesis theories are based upon the thermal Freeze-out mechanism. 
It is to be noted that the assumption behind the weakly interacting massive particle (WIMP) dark matter is that the coupling between the dark matter and SM particles cannot be too small; otherwise, it will never reach thermal equilibrium. Similarly, we cannot have large coupling for a small dark matter mass, and it will violate the perturbative-unitarity and the direct detection limits~\cite{Aprile:2018dbl}.  So far, we do not have any signature in those direct detection experiments~\cite{Aprile:2018dbl}.
Moreover, most of the single component WIMP dark matter model~\cite{Burgess:2000yq,Deshpande:1977rw,Khan:2016sxm,Chaudhuri:2015pna} are tightly constrained by the recent direct detection limits~\cite{Aprile:2018dbl}. People have also introduced the multi-component of the dark matter models~\cite{Dienes:2011ja} to avoid these limits.
Recently the authors of the Ref.~\cite{Hall:2009bx}, have suggested the idea of a Freeze-in mechanism where the dark matter candidate interacts with the other particle feebly, called Feebly Interacting Massive Particles (FIMPs).
It is assumed that, initially the dark matter density remains tiny in the early universe and get populated through the decay and/or annihilation of the other heavy particles. 
We found manly two types of Freeze-in mechanisms in the literature. 
The infra-red (IR) Freeze-in~\cite{Borah:2018gjk,Hall:2009bx,Biswas:2016bfo}, which is based on renormalizable theory
while the ultra-violet (UV) Freeze-in~\cite{Elahi:2014fsa} model consists of higher dimension operators, and the relic density in the UV Freeze-in mechanism depends on the reheat temperature.

The lepton flavour violating (LFV) processes ($\mu \rightarrow e \gamma$) along with the electron and muon anomalous magnetic moment are also a striking indication of the BSM frameworks. Recently, the muon g-2 collaboration at Fermilab has reported an staggering measurement of
the anomalous magnetic dipole moment of $\mu^{\pm}$ with a 3.3$\sigma$ deviation from the SM prediction
achieving a combined experimental average of ~\cite{Abi:2021gix, Bennett:2006fi, Parker:2018vye}:
$\delta a_{\mu}=a_{\mu}^\text{exp}-a_{\mu}^\text{SM}=(2.51\pm 0.59)\times 10^{-9}$.
This gives new hope in the BSM physics for a long-standing tension between SM and experimental data that was previously reported by the E821 experiment at Brookhaven National Laboratory \cite{Muong-2:2006rrc}, and it gets serious attention within a short period \cite{Borah:2021jzu,Dasgupta:2021dnl,Cen:2021iwv,Yang:2021duj,Das:2021zea,Escribano:2021css,Athron:2021iuf,Kowalska:2017iqv, Calibbi:2018rzv,Kowalska:2020zve,Kawamura:2020qxo,Baker:2021yli}.
{To explain these data, one can also alternatively use the lattice QCD, the hadronic vacuum polarization contribution to the muon anomalous magnetic moment based on gauge ensembles with $N_f = 2 + 1$ flavours of $O(a)$ improved Wilson quarks~\cite{Gerardin:2019rua,Gerardin:2020gpp,Chao:2021tvp,FermilabLattice:2019ugu}.}

In this work we add a $Z_2$-odd real scalar Higgs triplet with hypercharge $Y=0$ to the SM.
This model is popularly termed as the hyperchargeless Higgs
triplet model in the literature~\cite{Blank:1997qa}.
Along with this Higgs triplet, two additional vector-like $SU(2)_L$ doublets\footnote{In this model, we need two generations of fermion doublet to address all the light neutrino parameters \cite{Das:2020hpd}} and neutral singlet fermions are also added in this model (to cancel the effect of the anomaly), which can help to explain the dark matter as well as all the light neutrino parameters. In a parallel footing, we also probe the couplings and other models parameters to the latest results on muon anomalous magnetic moment and check whether this model can accommodate those results or not.
The Higgs triplet~\cite{Khan:2016sxm} alone can produce the relic density for dark matter mass $M_{DM}>2.3$ TeV onward but, it fails to provide the exact relic density since the co-annihilation channels dominate at the low mass region. Similarly, the extension of $Z_2$-odd vector-like neutral fermion could explain the DM, however, most of the parameter space ruled out by the present direct detection experiments. { It is worthy mentioning that the region $M_H +M_{H^\pm}<M_W$ is ruled out from the decay of $W$-boson. Though, in the region $2 M_H, 2M_{H^\pm}<M_h$, there still have some allowed parameters even from the 14 TeV LHC. In the only HTM (i.e., without vector-like fermion) model, the Higgs-125 GeV invisible decay width put only strong bounds in the region $2 M_H, 2M_{H^\pm}<M_h$. Interestingly, the $H^\pm$ could be long-lived in the HTM model as the mass is almost degenerate, the decay time we found $\sim 10^7$ sec. However, in the presence of vector-like fermion $H^\pm$ could decay quickly.}
We find that Higgs triplet and vector-like fermions in this model can explain a large region of the dark matter parameter spaces including the neutrino low energy variables, allowed by all theoretical and experimental constants. To the best of our knowledge, this model has not discussed in the literature, which motivates us to go for a detailed analysis.

The paper is organized as follows. Section~\ref{model} starts with a detailed description of the model. We revisit the detailed constraints for the extended HTM model in Sec.~\ref{sec:constraints}.
The new constraints from the LFV decay and contributions to the muon anomalous magnetic moment are discussed in the Sec.~\ref{sec:lfvanm}. We discuss the neutrino mass and mixing angle in Sec.~\ref{sec:nmass}. The FIMP dark matter parameter spaces are shown in detail in Sec.~\ref{sec:DM}.
Finally, we conclude our work in Sec.~\ref{sec:conclu}.
\section{Model}\label{model}
We extended the SM particle content by a triplet $T$, a vector-like fermion (VLF) singlet $N_S$ and two VLF doublet
 $F_{D,i}=(N_D~~E_D^-)^T_{i=1,2}$. The choice of a vector-like particle in this model will keep the model anomaly free \cite{Pal:1690642}.
We impose a discrete $Z_2$ symmetry on this model such that the scalar triplet and vector-like fermions are $odd$ ($\Psi_{BSM}\rightarrow-\Psi_{BSM}$), whereas the SM fields are $even$ ($\Psi_{SM}\rightarrow \Psi_{SM}$) under this transformation.
The extra scalar triplet consists of a pair of singly-charged fields $H^\pm$ and a $CP$-even neutral scalar field $H$.
The doublet and triplet scalar are conventionally written as~\cite{Khan:2016sxm, Chen:2008jg, Fiaschi:2018rky}
\begin{eqnarray}
\Phi=\left(
\begin{array}{c}
{G_{1}^+} \\
\frac{1}{\sqrt{2}}(v_1+{h^0}+i{G^0})
\end{array}
\right)\, ,
\qquad \qquad
T=\left(
\begin{array}{c}
{H^+}\\
{H}\\
-{H^-}
\end{array}
\right)\, \equiv \left(
\begin{array}{cc}
\frac{H}{\sqrt{2}} & H^+\\
H^- & -\frac{H}{\sqrt{2}}
\end{array}
\right).\label{Tripfield1}
\end{eqnarray}

The kinetic part of the Lagrangian is given by,
\begin{equation}
{\cal L}_k=\mid D_\mu \Phi \mid^2+\frac{1}{2}\mid D_\mu T\mid^2 \, ,
\end{equation}
where, the covariant derivatives are defined as,
\beq
D_\mu \Phi = \biggl( \partial_\mu +i \frac{g_2}{ 2}
 \sigma^aW^a + i\frac{g_1}{ 2} Y B_\mu\biggr)\Phi
~~~\text{and}~~~
D_\mu T = \biggl(\partial_\mu +i g_2 t_aW^a\biggr)T\, ,
\eeq
here, $W^{a}_\mu$ ($a$=1,2,3) are the ${ SU(2)_L}$ gauge bosons, corresponding to three generators of ${ SU(2)_L}$ group and $B_\mu$ is the ${ U(1)}_Y$ gauge boson.
$\sigma^a ~(a=1,2,3)$ are the Pauli matrices,
and $t_a$ can be written as follows
\begin{equation}
t_1=\frac{1}{ \sqrt{2}}
\left(
\begin{array}{ccc}
0&1&0\\
1&0&1\\
0&1&0
\end{array}\right),\quad
t_2=\frac{1}{\sqrt{2}}
\left(\begin{array}{ccc}
0&-i&0\\
i&0&-i\\
0&i&0
\end{array}\right),\quad
t_3=\left(\begin{array}{ccc}
1&0&0\\
0&0&0\\
0&0&-1
\end{array}\right)\, .
\end{equation}

The tree-level scalar potential with the Higgs doublet and the real scalar triplet is invariant under $SU(2)_L\times U(1)_Y$ transformation. This is given by,
\begin{eqnarray}
V(\Phi,T)&=&\mu_1^2\mid \Phi\mid^2
+\frac{\mu_2^2}{ 2}\mid T\mid^2
+{\lambda_1} \mid \Phi\mid^4 
+\frac{\lambda_2}{ 4}\mid T\mid^4
+\frac{\lambda_3}{ 2}\mid \Phi\mid^2\mid T \mid^2.
\label{ScalarpotTrip}
\end{eqnarray}
The scalar fields of the triplet do not mix with the scalar fields of SM doublet. 
After the EWSB, the scalar potential in Eq.~\ref{ScalarpotTrip} is then given by
\bea
V(h, H,H^\pm) &=&  \frac{1}{4} \left[ 2 \mu_1^2 (h+v)^2 + \lambda_1 (h+v)^4 +2 \mu_2^2 (H^2+2 H^+ H^-) \right. \nn \\
&& \left. + \lambda_2 (H^2 + 2 H^+ H^-)^2  + \lambda_3 (h+v)^2 (H^2+2 H^+ H^-) \right].
\eea
Here, $v \equiv v_{SM}$ and the mass of these scalar fields 
 $h$, $H$ and $H^\pm$ are given by
\bea
M_{h}^2 &=&  2 \lambda_1 v^2,\nn \\
M_{H}^2 &=& \mu_2^2 +  \frac{\lambda_3}{2} v^2, \\ 
M_{H^\pm}^2 &=& \mu_2^2 + \frac{\lambda_3}{2}  v^2  \,\nn .\label{MassTrip}
\eea
At tree-level, the mass of the neutral scalar $H$ and the charged particles $H^\pm$ are degenerate. If we include one-loop radiative correction, the charged particles become slightly heavier~\cite{Cirelli:2009uv,Cirelli:2005uq} than the neutral ones. The mass difference between them is given by,

\beq
\Delta M=(M_{H^\pm}-M_{H})_{1\text{-}loop}=\frac{\alpha M_{H}}{4\pi}\Big[f\Big(\frac{M_W}{M_{H}}\Big) -c_W^2 f\Big(\frac{M_Z}{M_{H}}\Big)\Big]
\label{massdifftrip},
\eeq
with,
$f(x)=-\frac{x}{4}\Big\{ 2 x^3 ~ {\rm log}(x)+(x^2-4)^{\frac{3}{2}}~ {\rm log}\left( \frac{x^2-2-x\sqrt{x^2-4}}{2} \right)\Big\}$ and $c_W$ being the $cosine$ of the Weinberg angle.
It has been shown in Refs~\cite{Cirelli:2009uv,Cirelli:2005uq} that the mass splitting between charged and neutral scalars remains $\sim150$ MeV for $M_H = 0.1-5$ TeV.

The Lagrangian for the new fermionic interaction are given by, 
\begin{eqnarray}
\nn \mathcal{L_F}&=&\overline{F}_{D,i}\gamma^{\mu}D_{\mu}F_{D,i}+\overline{N}_S\gamma^{\mu}D_{\mu}  {N}_S-M_{ND}^{ik} \overline{F}_{D,i} F_{D,k} - M_{NS}\overline{N}_S {N}_S,\\
\mathcal{L}_{int}&=&-Y_{N,i}\overline{F}_{D,i} \phi  {N}_S - Y_{fli} \overline{L}_l F_{D,i} \, T + h.c.~,\label{lint}
\end{eqnarray}
here, $D_{\mu}$ stands for the corresponding covariant derivative of the doublet and singlet fermions. The indices expand as $l=e,\mu,\tau$ and $i=1,2$.
The mass matrix for these neutral fermion fields is given by,
\begin{eqnarray}
\mathcal{M}=\begin{pmatrix}
M_{NS}&M_X^1 & M_X^2\\
M_X^{1,\dagger}&M_{ND}^{11} & M_{ND}^{12}\\
M_X^{2,\dagger}& M_{ND}^{21} & M_{ND}^{22}\\
\end{pmatrix},
\label{eq:mass}
\end{eqnarray}
where, $M_X^1=\frac{v \, Y_{N,1} }{\sqrt{2}}$ and $M_X^2=\frac{v \, Y_{N,2} \,}{\sqrt{2}}$. The neutral component of the fermion doublets ($N_{D,1}$ and $N_{D,2}$) and the singlet charged fermion ($ {N}_S$) mix at tree level. For simplicity, Let us assume the second VLF doublet is decoupled, i.e., $Y_{N,2}=0$ and $M_{ND}^{21}=M_{ND}^{12}=0$~\footnote{We discuss the diagonalization of these fermions in Appendix~\ref{app:A} for the non-zero of-diagonal components in Eq.~\ref{eq:mass}}. The mass eigenstates are obtained by diagonalizing the mass matrix with
a rotation of the $ ({N}_S$  $N_{D,1}$ $N_{D,2})$ basis as,
\begin{eqnarray}
\begin{pmatrix}
\chi_1 \\\chi_2\\ \chi_3\\
\end{pmatrix}=\begin{pmatrix}
\cos\beta&\sin\beta & 0\\
-\sin\beta&\cos\beta &0\\
0 & 0 & 1\\
\end{pmatrix}\begin{pmatrix}
N_S\\ N_{D,1} \\ N_{D,2}\\
\end{pmatrix}, {~\rm with~} \tan 2 \beta = \frac{2 M_X^1}{M_{ND}^{11}-M_{NS}}
\end{eqnarray}
Diagonalization of Eq.~\ref{eq:mass} gives the following eigenvalues for the new leptons ($M_{ND}^{11}-M_{NS} \gg M_X^1$) as,
\begin{eqnarray}
&M_{\chi_1} = M_{NS} - \frac{2 (M_X^1)^2}{M_{ND}^{11}-M_{NS}},\,
M_{\chi_2} = M_{ND}^{11} + \frac{2 (M_X^1)^2}{M_{ND}^{11}-M_{NS}}, \, M_{E_1^\pm} =M_{ND}^{11}, \,
M_{E_2^\pm}=M_{\chi_3} = M_{ND}^{22},\nn
\end{eqnarray}
$E_1^\pm \equiv E_{D,1}^\pm$ and $E_2^\pm \equiv E_{D,2}^\pm$ are the charged component  fermion fields. { The fermion state $\chi_1$ can serve as a dark matter candidate for the choice of mass $M_{\chi_1} < M_{\chi_{2,3}, E_{1,2}^\pm, H,H^\pm}$. With this choice of parameters $Y_{N,2}=0$ and $M_{ND}^{21}=M_{ND}^{12}=0$, the second doublet becomes decoupled from the first doublet and singlet vector-like fermion. Hence, it does not affect the dark matter density calculations; however, it is important to get all the neutrino low energy variables. In the absence of the second doublet, we can not explain one of the neutrino parameters: mass differences or mixing angles. We will provide a detailed discussion on the new region of the allowed parameter spaces and the effect of the presence of additional $Z_2$-odd fermion in the sections of  dark matter~\ref{sec:DM} and neutrino mass \ref{sec:nmass}.}

The parameter space of this model is constrained by various bounds arising from theoretical considerations like absolute vacuum stability and unitarity of the scattering matrix, observation phenomenons like dark matter relic density. 
The direct search limits at LEP and electroweak precision measurements also put severe restrictions on the model.
The recent measurements of the Higgs invisible decay width and signal strength at the LHC put additional constraints.
The dark matter (DM) requirement saturates the DM relic density all alone restricts the allowed parameter space considerably.
\section{Constraints on the model}
\label{sec:constraints}
\subsection{Vacuum stability bounds}
The condition for the absolute stability of the vacuum comes from requiring that the scalar potential is bounded from below, i.e., it should not approach negative infinity along any direction of the field space for large field values~\cite{Khan:2016sxm}. 
One can use the copositivity criteria~\cite{Kannike:2012pe} and calculate the required conditions for the absolute stability/bounded from below the scalar potential. The tree-level scalar potential is absolutely stable if,
\beq
\lambda_{1}(\Lambda) \geq 0, \quad  \lambda_{2}(\Lambda) \geq 0, \quad  \lambda_{3}(\Lambda) \geq - 2~\sqrt{ \lambda_{1}(\Lambda)\lambda_{2}(\Lambda)}. \label{stabilitybound}
\eeq
The coupling constants are evaluated at a scale $\Lambda$ using RGEs.
\subsection{Perturbativity and Unitarity bounds}
\label{ssec:perturbativity}
To ensure that the radiatively improved scalar potential remains perturbative at any energy scale ($\Lambda$), one must put the following conditions,
\beq
\mid \lambda_{1,2,3}\mid \lesssim 4 \pi ~~~{\rm and}~~~\Big |\frac{\lambda_4}{\Lambda}\Big | \lesssim 4 \pi.
\eeq
The most stringent unitary bound comes from the sub-matrix ${\cal M}_3$ (see Ref.~\cite{Khan:2016sxm}) corresponds to scattering fields
$(G_1^+~G_1^-,~\eta^+~\eta^-,$ $~\frac{G^0~G^0}{\sqrt{2}},~\frac{h^0~h^0}{\sqrt{2}},$ and $\frac{\eta^0~\eta^0}{\sqrt{2}})$ and it is given by,
\beq
{\cal M}_3=
\begin{pmatrix}
 4 \lambda_1 & \lambda_3 &  \sqrt{2} \lambda_1 & \sqrt{2}\lambda_1 & \frac{\lambda_3}{ \sqrt{2}} \\
 \lambda_3 & 4 \lambda_2 & \frac{\lambda_3}{ \sqrt{2}} & \frac{\lambda_3}{ \sqrt{2}} &  \sqrt{2} \lambda_2 \\
 \sqrt{2} {\lambda_1} & \frac{\lambda_3}{ \sqrt{2}} & 3 \lambda_1 & \lambda_1 & \frac{\lambda_3}{2} \\
 \sqrt{2} \lambda_1 & \frac{\lambda_3}{ \sqrt{2}} & \lambda_1 & 3 \lambda_1 & \frac{\lambda_3}{2} \\
 \frac{\lambda_3}{ \sqrt{2}} & \sqrt{2} \lambda_2 & \frac{\lambda_3}{2} & \frac{\lambda_3}{2} & 3 \lambda_2
\end{pmatrix}.
\eeq
Unitary constraints of the scattering processes demand that the eigenvalues $ 2\lambda_1,~2 \lambda_1,~2 \lambda_2,$ and $\frac{1}{2} \left(6 \lambda_1+5 \lambda_2\pm\sqrt{(6 \lambda_1-5 \lambda_2)^2+12 \lambda_3^2}\right)$ of the matrix should be less than $8\pi$~\cite{Khan:2016sxm}.
\subsection{Bounds from electroweak precision experiments}
Electroweak precision data has imposed severe bounds on new physics models via Peskin-Takeuchi $S, ~T, ~U$ parameters~\cite{Peskin:1991sw}. The additional contributions from the scalar triplet of this model are given by~\cite{Forshaw:2003kh, Forshaw:2001xq},
{\allowdisplaybreaks
\bea
S  & \simeq & 0, \\
T   &=&  \frac{1}{8\pi} \, \frac{1}{\sin^2\theta_W \cos^2\theta_W} \left[ \frac{M^2_H + M^{2}_{H^\pm}}{M^{2}_{Z}} \;
- \; \frac{2  M^{2}_{H^\pm} M^2_H}{M^{2}_{Z}(M^2_H -  M^{2}_{H^\pm})} \log\left(\frac{M^2_H}{M^{2}_{H^\pm}}\right)
\right] \nonumber \\
 &\simeq & \frac{1}{6\pi} \, \frac{1}{\sin^2\theta_W \cos^2\theta_W} \; \frac{(\Delta M)^{2}}{M^{2}_{Z}}, \\
U   &=&  -\frac{1}{3 \pi} \left( M^4_H\log \left( \frac{M^2_H}{M^{2}_{H^\pm}} \right) \frac{ (3 M^{2}_{H^\pm}-M^2_H)}{(M^2_H-M^{2}_{H^\pm})^3} + \frac{5(M^4_H+M^{4}_{H^\pm})-22 M^{2}_{H^\pm} M^2_H}{6(M^2_H-M^{2}_{H^\pm})^2} \right) \nonumber \\ 
&\simeq &  \frac{\Delta M}{3 \pi M_{H^\pm}},
\eea}
where, $\Delta M = M_{H^\pm} - M_{H}\approx 160$ MeV~\cite{Khan:2016sxm}.
$M_{H^\pm}$ and $ M_H$ are almost degenerate for $M_{H^\pm,H} > M_h$. The contributions to the $T~{\rm and}~U$ parameters from this model are also neglected~\cite{Khan:2016sxm, Baak:2014ora}.
{ Similarly the contribution from the $SU(2)$ singlet neutral fermion is zero whereas in the absence of mixing and $M_{N_{D,i}},M_{E_{D,i}^\pm}>>\Delta M_{D,i}=M_{E_{D,i}^\pm}-M_{N_{D,i}} (i=1,2)$, the contribution from the doublet fermion doublet are given by~\cite{Peskin:1991sw,Hieu:2020hti},

\bea
S  & \simeq & \frac{2}{3 \pi} \, \sum_{i=1,2} \frac{\Delta M_{D,i}}{M_{N_{D,i}}}, \\
T   &\simeq&  \frac{1}{6\pi}  \frac{1}{\sin^2\theta_W} \,\sum_{i=1,2} \frac{(\Delta M_{D,i})^{2}}{M^{2}_{W}} \\
U   &\simeq&  \frac{11}{30\pi} \, \sum_{i=1,2} \frac{(\Delta M_{D,i})^{2}}{M^{2}_{W}},
\eea
In the tree-level, $\Delta M_{D,i}=0$ and the loop-level separation is also small, hence the contributions to the $S, T$ and $U$ parameters from these additional fermion are almost negligible. 
}
\subsection{LHC diphoton signal strength constraints}
\label{diphoton}
Gluon fusion at LHC gives the dominant production cross-section of $h$, and the Higgs to diphoton signal strength $\mu_{\gamma\gamma}$ in this extended HTM model can be written as,
\beq
\mu_{\gamma\gamma} = \frac{\sigma(gg\ra h\ra\gamma\gamma)_{EHTM}}{\sigma(gg\ra h\ra\gamma\gamma)_{\rm SM}}= \frac{\sigma(gg\ra h)_{EHTM}}{\sigma(gg\ra h)_{SM}}  \frac{Br(h \rightarrow {\gamma\gamma})_{\rm EHTM}}{Br(h \rightarrow {\gamma\gamma})_{\rm SM}}.
\eeq
We use the narrow width approximation as $\Gamma_h^{total}/{M_h} \ra 0$. The $\mu_{\gamma\gamma}$ can be simplified as,
\beq
\mu_{\gamma\gamma} = ~\frac{\Gamma^{total}_{h,\rm SM}}{\Gamma^{total}_{h,\rm EHTM}}~\frac{\Gamma(h\rightarrow \gamma\gamma)_{\rm EHTM}}{\Gamma(h\rightarrow \gamma\gamma)_{\rm SM}}.\label{Hdecay}
\eeq
 Since, the couplings $h\rightarrow f\bar{f}/ VV$, ($V$ stands for vector bosons) are same as in SM.
 
The new charged fermions $E_{1,2}^\pm$ and charged Higgs $H^\pm$ in this model will alter the decay width of $h\rightarrow\gamma\gamma $, $ Z\gamma$ through one-loop. 
At the same time, if the mass of the extra scalar particles ($\chi_{1,2,3}, E_{1,2}^\pm, H,H^\pm$) happen to be lighter than the half of the Higgs mass $M_h/2$, then they
might also contribute to the invisible decay of the Higgs boson.
We are using the global fit analysis~\cite{Belanger:2013xza}, that such an invisible branching ratio is less than $\sim 20 \%$. In Eq.~\ref{Hdecay}, the first ratio provides a suppression of $\sim 0.8-1$.
For $M_{\chi_{1,2,3}, E_{1,2}^\pm, H,H^\pm}> M_h/2$, the ratio becomes $\frac{\Gamma^{total}_{h,\rm SM}}{\Gamma^{total}_{h,\rm EHTM}}\approx 1$. Therefore, the modified Higgs to diphoton signal strength, $\mu_{\gamma\gamma}$ can be written as,
\beq 
 \mu_{\gamma\gamma} \approx \frac{\Gamma(h\rightarrow \gamma\gamma)_{\rm EHTM}}{\Gamma(h\rightarrow \gamma\gamma)_{\rm SM}}\, .
 \label{mugagahighIT}
\eeq
{ In this model, the additional contribution arising due to $E_{12}^\pm$ and $H^\pm$ are incorporated in the total decay width as given by~\cite{Djouadi:2005gj}, }
\bea
\Gamma(h\rightarrow \gamma\gamma)_{\rm EHTM}&=&\frac{\alpha^2 M_h^3}{ 256\pi^3
v^2}\Big|\sum_{f}N^c_fQ_f^2y_f
F_{1/2}(\tau_f)+ y_W F_1(\tau_W) + Q_{E_1^\pm}^2 y_{hE_1^+ E_1^-}
F_{1/2}(\tau_{E_1^\pm})\nn\\
&&~~~~~~~~~~~~~~ + Q_{E_2^\pm}^2 y_{hE_2^+ E_2^-}
F_{1/2}(\tau_{E_2^\pm}) +Q_{H^{\pm}}^2\frac{v\mu_{{hH^+H^-}}}{
2M_{H^{\pm}}^2}F_0(\tau_{H^{\pm}})\Big|^2
\label{hgaga},
\eea
where $\tau_i=M_h^2/4M_i^2$. $Q_{X}$ ($X=f,{E_1^\pm},{E_2^\pm},H^{\pm}$) denotes electric charges of corresponding particles. $N_f^c$ is the color factor. $y_f$ and
$y_W$ denote the Higgs couplings to $f\bar{f}$ and $W^+W^-$. $\mu_{hH^+H^-}=\lambda_3  v_{SM} $ stands for the coupling constant of $hH^+H^-$ vertex and $y_{hE_{1,2}^+ E_{1,2}^-}=0 $ as $E_{1,2}^\pm$ are $Z_2$ odd. The loop functions $F_{(0,\,1/2,\,1)}$  can be found in Ref~\cite{Djouadi:2005gj}.

Recently, the diphoton rates $\mu_{\gamma\gamma}$ of the observed Higgs to the SM prediction are measured by the ATLAS~\cite{Aad:2014eha,ATLAS:2019cid} and CMS~\cite{Khachatryan:2014ira,CMS:2018uag} collaborations. At present the combined result from these experiments of $\mu_{\gamma\gamma}$ is $1.14^{+0.19}_{-0.18}$~\cite{Khachatryan:2016vau}. 

{
In $\Gamma(h\rightarrow \gamma\gamma)_{\rm EHTM}$ (see Eq.~\ref{hgaga}), a positive $\lambda_3$ leads to a destructive interference between Higgs triplet and the SM contributions and {\it vice versa}.
One can see from the Eq.~\ref{hgaga}, the contribution to the Higgs diphoton channel is proportional to $\frac{\lambda_3}{M_{H^\pm}^2}$.
If the charged scalar mass is greater than $300$ GeV, then the contributions of $H^\pm$ to the diphoton signal is negligibly small.
}
\section{Lepton flavour violating decay and anomalous magnetic moments}
\label{sec:lfvanm}
\begin{figure}[h]
	\centering
	\includegraphics[scale=0.70]{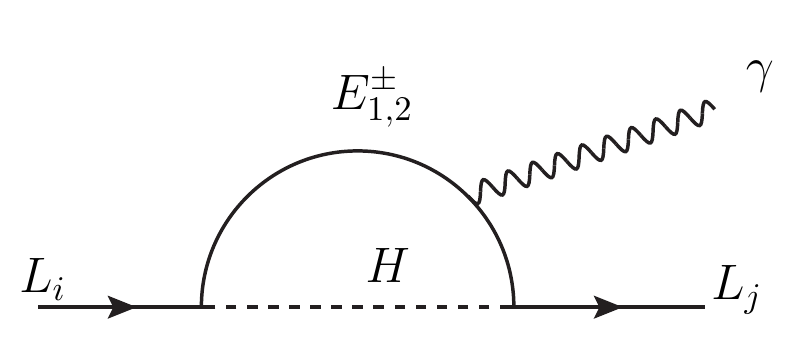}
	\hspace{0.2cm}
		\includegraphics[scale=0.70]{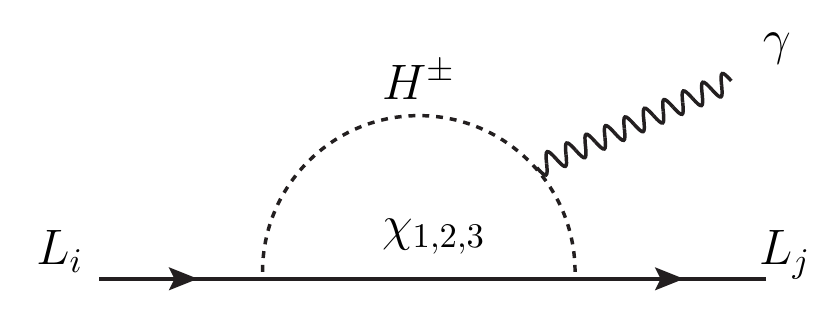}
	\caption{The diagrams contribute to LFV and both electron and muon anomalous magnetic moments.}\label{Fig:lfv}
\end{figure}

Among the different lepton number violating processes, the radiative (through one-loop) muon decay
$\Gamma(\mu\rightarrow e \gamma)$ is one of the popular and tightly constraints. In this present model, the diagram is mediated by charged particles $E_1^\pm, E_2^\pm, H^{\pm}$ and $\chi_1, \chi_2, \chi_3, H$ present in the internal lines of the one-loop diagrams~\ref{Fig:lfv}.
The mixing angle between the neutral component of singlet fermion with other neutral component of doublet fermions are very small, hence the contributions through $\chi_1$ is neglected.
The corresponding expression for the branching ratio is given by~\cite{Akeroyd:2009nu},
\begin{eqnarray}
	{\rm BR} (\mu\rightarrow e \gamma) \approx \sum_{i=1}^2 \frac{3 \alpha_{em}}{32 \pi G_F^2} \Big | Y_{f\mu i} Y_{fe i} \frac{ F(M_{E_{i}^\pm}^2/M_H^2) }{ M_H^2} \Big |^2 ,
	\label{eq:fl}
\end{eqnarray}
where, $
F(x)=\frac{x^3-6 x^2 +3 x+ 2 + + x ln(x)}{6 \, (x-1)^4}
$.
The most recent experimental bounds for LFV could be found in Ref.~\cite{Baldini:2018nnn} is given by $ {\rm BR} (\mu\rightarrow e \gamma) < 4.2 \times 10^{-13}$ at $90\%$ CL. We need small Yukawa coupling $Y_{f\mu i} $ and/or $Y_{fei}$ to get the allowed parameter spaces in a desirable range. It is noted that these Yukawa couplings can also alter the neutrino mass and mixing angle (will be discussed in the next section) and the muon as well as electron anomalous magnetic moments. The numerical expression for additional contribution to the electron ($l=e$) and muon ($l=\mu$) anomalous magnetic moments in this model is expressed as~\cite{Chao:2008iw,Chen:2019nud},
\begin{eqnarray}
\Delta \alpha_i &=& \frac{ Re[Y_{fli}^2] m_l^2 }{8 \pi^2}  \int_0^1 \, \frac{ x (1-x)^2}{(x-x^2) m_l^2  + (x-1) M_{E_i^\pm}^2 - x M_{H}^2 } \nn\\
 & & + \, \frac{ Re[Y_{fli}^2] m_l^2 }{16 \pi^2}  \int_0^1 \, \frac{ x (1-x)^2}{(x^2-x) m_l^2  + x  M_{E_i^\pm}^2 - (x-1) M_{H^\pm}^2 }.
\end{eqnarray}
This model shows that the muon decay $\Gamma(\mu\rightarrow e \gamma)$ and neutrino mass and mixing angles put tighter constraints on the model parameter spaces. We find that the contributions to both anomalous magnetic moments are almost negligible in this model. Interestingly, one can keep a large $Y_{f\mu i} (i=1,2)$ to achieve the discrepancy between the SM predictions
$\delta a_{\mu}=a_{\mu}^\text{exp}-a_{\mu}^\text{SM}=(2.51\pm 0.59)\times 10^{-9}$ ~\cite{Abi:2021gix}. In this case, we choose to keep a very small $Y_{fe i} (i=1,2)$ to satisfy ${\rm BR} (\mu\rightarrow e \gamma) < 4.2 \times 10^{-13}$. With these choices of parameters, one cannot get all the neutrino low energy variables as measured by the experiments. In the next section we will discuss the neutrino oscillation parameters for this anomaly.
\section{Neutrino Mass and mixing angles}
\label{sec:nmass}
\begin{figure}[h]
	\centering
	\captionsetup{justification=centering}
	\includegraphics[scale=1.20]{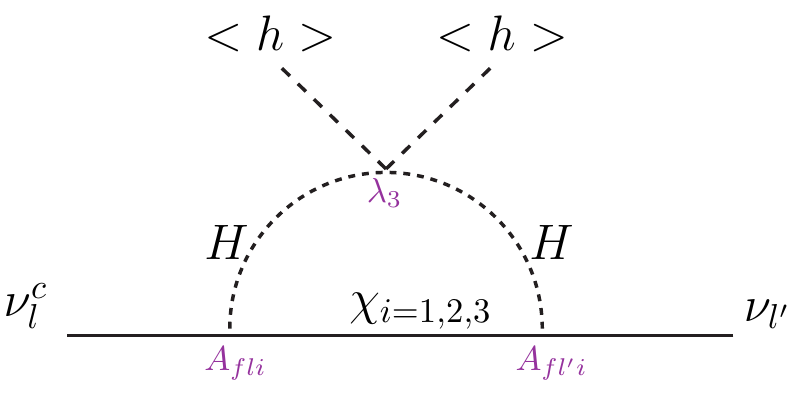}
	\caption{ One loop diagram for neutrino mass generation.}\label{fig:numass}
\end{figure}

In this study, we will provide a brief overview of the neutrino mass generation at the one-loop level. 
 The second term of the interaction Lagrangian from Eq. \eqref{lint} is responsible for the neutrino mass generation via 1-loop level, shown in Fig. \ref{fig:numass}. The Yukawa interaction associated with the interaction term, $Y_{fli}$ does violates lepton number conservation in this model. The neutral component of the fermion doublets ($N_{D,i}$) and singlet fermion ($N_S$) mixed at tree level via the mixing angle $\beta$. Hence, the Yukawa interaction also modified accordingly. One can find the modified Yukawa interaction $A_{fli}=\cos\beta~ Y_{fli}$ involved in the neutrino mass generation process. 
The neutral $Z_2$-odd vector-like fermions and scalar involved in the radiative neutrino mass generation after the EWSB are shown in Fig.~\ref{fig:numass}.
Summing over all the two-point function contributions, we arrive at the neutrino mass matrix component as~\cite{FileviezPerez:2009ud},
\begin{eqnarray}
(M_{\nu})_{ll'}=\sum_{i,l,l'}\frac{A_{fli}^\dagger A_{fl'i}}{32 \, \pi^2} \, (  \lambda_3 v^2 )  \, \mathcal{I} (M_{\chi_i}, M_{H}),  
\label{eq:n1}
\end{eqnarray}
where, $\lambda_3$ is the Higgs portal quartic coupling, and $v=264.221$ GeV is the SM vacuum expectation value. The indices $i=1,2,3$ stand for the three neutral fermion. $M_{\chi_{i=1,2,3}}$ are the masses for the neutral heavy fermions. The loop function $I (M_{\chi_i}, M_{DM}) $ is defined as~\cite{FileviezPerez:2009ud},
\begin{eqnarray}
\mathcal{I} (M_{\chi_i}, M_{H}) = 4 M_{\chi_i}  \frac{M_{H}^2 -M_{\chi_i}^2 + M_{\chi_i} \, log(\frac{M_{\chi_i}^2}{M_{H}^2})}{(M_{H}^2-M_{\chi_i}^2)^2}.
\label{eq:n2}
\end{eqnarray}

Neutrino mass eigenvalues are obtained by  diagonalizing the above mass matrix $(M_{\nu})_{ij}$ using the well established PMNS matrix, consists of three mixing angles ($\theta_{12},\theta_{13},\theta_{23}$) and three phases (one Dirac and two Majorana). The diagonal mass matrix can be written as $ m_{Diag}=U_{\rm PMNS}^\dagger M_\nu U_{\rm PMNS}$. Moreover, it is essential to ensure that the choice of Higgs portal quartic coupling $\lambda_3$, Yukawa couplings, as well as other parameters involved in the light neutrino mass matrix component..
From the above expressions (eqns.~\ref{eq:n1} and~\ref{eq:n2}), light neutrino masses and mixing angles can exactly be explained by adjusting the Higgs portal quartic coupling $\lambda_3$, Yukawa couplings and mass parameters present in equation \eqref{eq:n1}.
For a few hundred GeV, heavy scalar and neutral fermions, one can choose small $\lambda_3 \sim \mathcal{O}(10^{-8})$ to get the small neutrino masses. It is important to highlight that the neutrino mass and mixing angles are almost independent of the dark matter mass $M_{\chi_1}$ and tiny mixing angle $\beta$.

A tiny $\beta$ is essential to get the exact relic density through the freeze-in mechanism. We also find that the second generation of the fermion doublet is necessary to generate all the neutrino mass differences and mixing angles according to the experimental data. In the absence of this doublet, it is unable to produce one or more light neutrino variables. For example, we can generate mixing angles $\theta_{12}=32.7^{\circ}$, $\theta_{13}=8.4^{\circ}$, $\theta_{23}=44.71^{\circ}$ and mass differences $\Delta m_{21}^2=7.44\times10^{-5}$ eV$^2$ and $\Delta m_{31}^2=4.9\times10^{-4}$ eV$^2$ with phases $\alpha=\delta=45^{\circ}$. 
Although $\Delta m_{31}^2$ is within the present 3$\sigma$ bound yet, $\Delta m_{21}^2$ deviates from the actual range for this choice of the parameters. For the other option of the input parameters, we can get $\Delta m_{21}^2$ is within the present 3$\sigma$ bound yet, $\Delta m_{31}^2$ is deviating from the actual range for this choice of the parameters. Therefore, we added another doublet without affecting dark matter results and with the choice of tiny couplings we can explain all the light neutrino parameters within the latest bound.

\begin{figure}[h!]
	\includegraphics[scale=0.35]{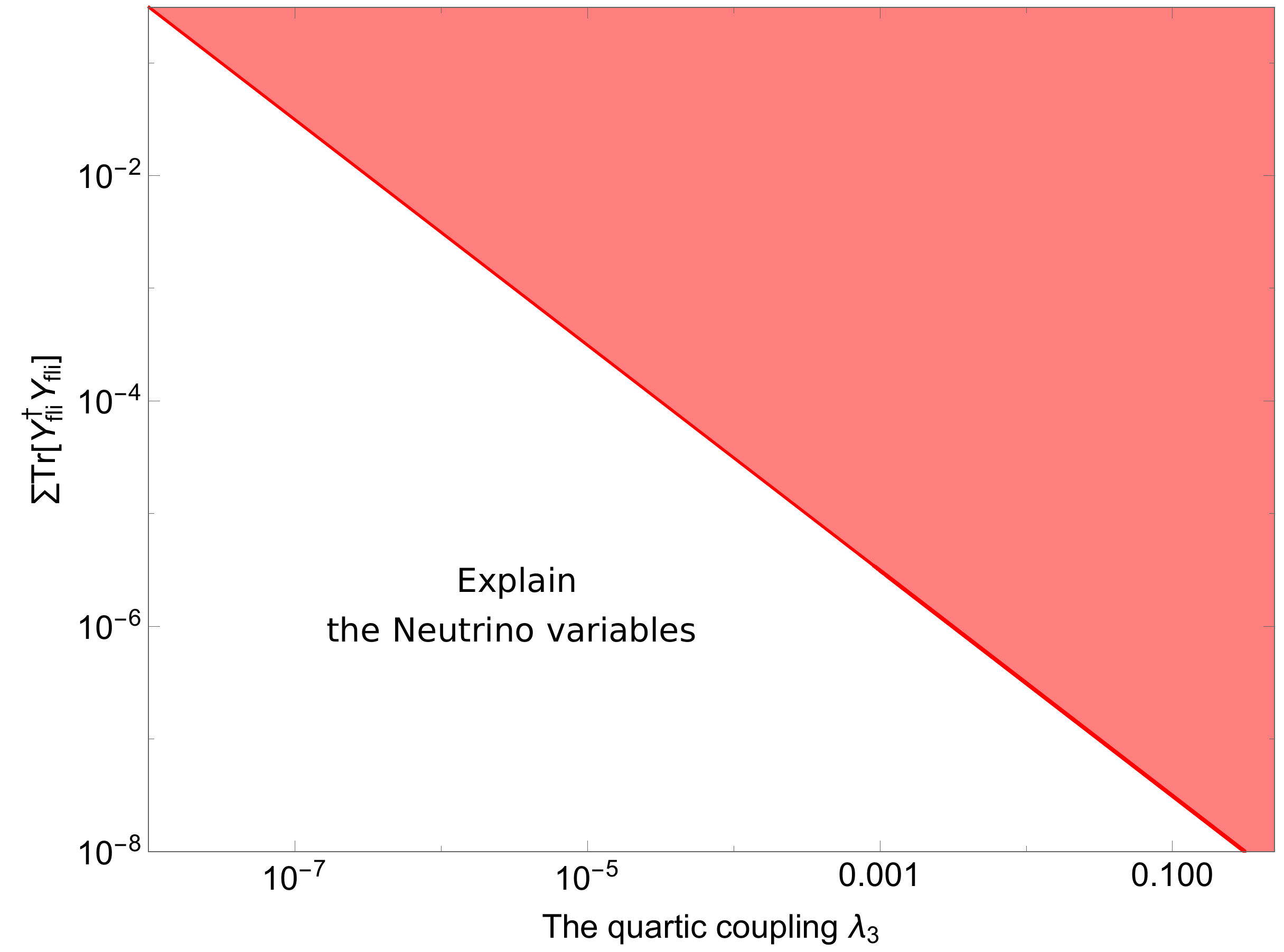}
	\caption{The plot stand for the neutrino low energy variable bounds in $\sum Tr|Y_{fli}^\dagger Y_{fli}|$ against the quartic coupling $\lambda_3$. We consider the fermion masses $M_{\chi_2}=1000$ GeV, $M_{\chi_3}=2500$ GeV and $M_{H}=1000$ GeV. The red region will violate one and/or more neutrino experimental parameters. }\label{neut}
\end{figure}
\begin{table}[h!]
\begin{center}\scalebox{0.8}{
\begin{tabular}{|c||c|c|c|}
\hline
~~~~~~Parameters~~~~~~ &\multicolumn{3}{c|}{Benchmark Points }\\
\cline{2-4} 
 & ~~~BM-I (NH) ~~~& ~~~BM-II (NH)~~~& ~~~BM-III (IH)~~~ \\
\hline
$M_{\chi_2}$~GeV&  $800$&$1000$&$880$\\
$M_{\chi_3}$~GeV&  $2500$&$2500$&$2000$\\
$M_{H}$~GeV&  $500$&$850$&$995$\\
$|Y_{fe1}|$  &   $0.00360362 $   &   $0.0141421 $  &  $0.0232594 $  \\
$|Y_{fe2}|$  &   $ 0.0462481 $   &    $0.0269258 $  &  $ 0.0282843 $  \\
$|Y_{f\mu1}|$  &   $0.000128141 $   &   $0.206155 $  &  $0.200679 $  \\
$|Y_{f\mu2}|$  &   $ 0.00420078 $   &    $0.400009 $  &  $ 0.632533 $  \\
$|Y_{f\tau1}|$  &   $0.00415482 $   &   $0.0613105 $  &  $0.0250098$  \\
$|Y_{f\tau2}|$  &   $ 0.0492443$   &    $ 0.0141421 $  &  $0.0150333 $  \\
$|\lambda_3|$  &   $ 6.6468 \times 10^{-8} $   &   $1.35 \times 10^{-9}  $  &  $ 3.657 \times 10^{-9}  $  \\
  \hline
\hline
Outputs&\multicolumn{3}{c|}{ Corresponding Low-energy variables}\\
\hline
 $\Delta m^2_{21}/10^{-5} ~{\rm eV^{2}}$&         7.552 &  7.735 & 2.5425 $\times 10^{2}$ (IH)\\
  $ \Delta m^2_{31}/10^{-3} ~{\rm eV^{2}}$&           2.504 & 2.46  & 7.46257 $\times 10^{-2}$ (IH)\\
         $ \theta_{12}$~~ rad&            0.5719  &   0.5708   & 0.5720\\
         $ \theta_{23}$~~ rad&            0.7803  &  0.7803   &   0.7803\\
         $\theta_{13}$~~ rad&            0.1469  &  0.1434  & 0.1469\\
         $\delta_{PMNS}$~~ rad&  $0.75$ &  $0.7$  &  $0.15$ \\
         $\alpha$~~ rad&         $0.5$  &  $0.6$  & $0.6$   \\
        $m_i$ ~eV&  $0.057634, 0.0298795, 0.0285878$& $0.0553511, 0.0260174, 0.0245639$  &  $0.0238301, 0.0550976, 0.0222092$\\
$Br(\mu \rightarrow e~\gamma) $ & $ 7.38 \times 10^{-19} $ &   $9.51 \times 10^{-15}$ &  $   3.06 \times 10^{-15}$ \\
\hline
\end{tabular}}
\end{center}
\caption{Three sets of benchmark points used in our analysis. These BPs are used to carry out our numerical analysis satisfying all the low energy constraints.}
\label{tab:neu}
\end{table}

Depending on the parameter space, here we can explain both the normal hierarchy (NH: $m_1<m_2<m_3$) and inverted hierarchy (IH: $m_3<m_1<m_2$). Three sets of such benchmark points and corresponding neutrino mass and mixing angles (within $3\sigma$) are displayed in table~\ref{tab:neu}. All three BPs are also allowed by the lepton flavor violation data. One can see that the masses of the heavy fermions and scalar particle are $\sim\mathcal{O}(1)$ TeV and $Y_{f \mu 2}\sim \mathcal{O}(10^{-2})$. 
	On the contrary, to achieve the precise value of the muon anomalous moment $\delta a_{\mu}=a_{\mu}^\text{exp}-a_{\mu}^\text{SM}=(2.51\pm 0.59)\times 10^{-9}$, one needs fine-tuning the Yukawa coupling $Y_{f\mu2}$ and this fine-tuning demands $Y_{f\mu2}>\mathcal{O}(1)$. Therefore, the region allowed by the muon $g-2$ data may also violate the unitary-perturbativity limits.
One can see from Fig. \ref{neut}, the current bounds in the light neutrino parameters are possible only with tiny Higgs portal coupling $\lambda_3$ and/or the Yukawa couplings $Y_{fli}$ for fermion masses $\mathcal{O}(1)$ TeV. The red region is the forbidden region for this model, as all the light neutrino parameters can not be explained at the same time in that region. Therefore, after discussing the model phenomenology and bounds achieved in the context of neutrino mass and mixing, we are now well equipped to carry out the dark matter and collider analysis. In the following section, we have studied FIMP dark matter in the context of a simplified version of this model with tiny coupling constants.

\section{Dark Matter}
\label{sec:DM}
We already discussed that one needs additional symmetry such that the dark matter could not decay and/or the decay lifetime is much larger than the lifetime of the universe. We have such $Z_2$-symmetry, which prohibits the lightest neutral $Z_2$-odd particle (dark matter) decay. In this model, we have total four such neutral $Z_2$-odd ($\chi_{1,2,3}$ and $H$) particles.
The eigenstate $\chi_1$ is the lightest neutral $Z_2$-odd fermion in this model and can be considered as a viable dark matter candidate for $M_{\chi_1}< M_H$. The other $Z_2$-odd neutral scalar field can also behave as a viable dark matter candidate for $ M_H < M_{\chi_1}$.
The co-annihilation channels ($\sigma (H^\pm H \rightarrow f_{SM} f_{SM})$) dominate in the relic density calculation for the $Z_2$-odd neutral scalar field through the Freeze-out mechanism. One can get the relic density for the scalar dark matter mass $M_H>2$ TeV, though the low mass region is completely ruled out.
On the other hand, for the fermionic case, we checked that various parameter space could give the exact relic density through the Freeze-out mechanism, however the large region of the parameter space is ruled out from the present direct detection experiment~\cite{Aprile:2018dbl}. Again, the low mass regions (except $\frac{M_h}{2} \pm 5$ GeV region) are forbidden from either the relic density and/or direct detection constraints.   
This study will mainly focus on the FIMP dark matter, i.e., the dark matter relic density achieved through the Freeze-in mechanism. Perhaps, the $Z_2$-odd neutral scalar field can also produce sufficient relic density, however, we do not discuss it here as this choice may not provide both normal as well as inverted hierarchy neutrino mass pattern in this model. Also, the region $M_{H}<\frac{M_{W,Z}}{2}$ are strictly prohibited from the invisible decay width constraints of the gauge bosons ($W,Z$).
Therefore, we only focus on the fermionic dark matter candidate for the mass region $keV-TeV$ in this study.

The central assumption is that the dark matter particle gets populated through the decay or annihilation of the heavy particles (SM and/or BSM) until the number density of the corresponding heavy particles species becomes Boltzmann-suppressed. In the existing literature~\cite{Borah:2018gjk,Hall:2009bx,Biswas:2016bfo}, it has been proved that if the same couplings are involved in both decay as well as scattering processes, then the former has the dominant contribution to DM relic density over the latter one. 
In this mechanism, it has been proved that the dark matter relic density depends only on the partial decay width (DM production channels only) of the mother particles~\cite{Hall:2009bx}. At the same time, the other decay channel (like, $\psi_{heavy}^{Mother}\rightarrow \psi_{light} \psi_{light} $) may reduce the mother particles density for the universe temperature temperature $T<M_{DM}$ rapidly.
However, for $T>M_{DM}$ the reverse processes ( $\psi_{light} \psi_{light} \rightarrow  \psi_{heavy}^{Mother}$) can compensate the mother particles density from the other bath particles.
Hence, the mother particle was thermally equilibrium with the other particles in the early universe~\cite{Hall:2009bx}.{
One can easily get ${\Gamma \over H(T) } \geq 1$ for all the mother particles which produce dark matter through decay and/or annihilation, where $\Gamma$ is the relevant decay width or annihilation rate for the production of mother particles from the annihilation of other particles and $H(T)$ is the Hubble parameter
$H(T) = \left( g^* \, \frac{\pi^2}{90} \, \frac{T^4}{\mpl^2} \right)^{1/2}$.
$\mpl=2.435\times 10^{18}$ GeV is the reduced Planck mass and $T$ is the temperature~\cite{Plehn:2017fdg,Hall:2009bx}.}
For the dark matter production channels, the forward as well backward processes are prolonged due to the smaller coupling strengths, hence the forward dark matter production processes are always dominant, and the reverse rate is almost zero during the expansion of the universe.
Therefore, the other decay and or annihilation channels of the mother particles will not affect the relic density calculations in the Freeze-in mechanism.

Thus, we have to solve only one Boltzmann equation for the evolution of the DM from various mother particles. In this case, we have to consider the sum of all DM production through the decay and annihilation channels of different mother particles.
It is noted that if the mother particles are not in thermal equilibrium in the early universe, then one must need to solve the evaluation of the mother particles and at the same time has to solve the evaluation for the DM particle~\cite{Biswas:2016bfo}.
For our case, the relic density can be calculated as~\cite{Plehn:2017fdg,Hall:2009bx,Biswas:2016bfo}
\bea
\Omega h^2 \approx 1.09 \times 10^{27} \, M_{ \chi_1 }  \,\sum_i  \frac{g_{\psi_{Heavy,i}} \, \Gamma_{\psi_{Heavy,i}} }{ M_{\psi_{Heavy,i}}^2} ,
\label{eq:totalOmega}
\eea
where, $\Gamma_{\psi_{Heavy,i}}$ is the partial decay width of various dark matter production channels. In this model, we find that the annihilation contributions are tiny as compared to the decay channels. 
The neutral component of the first vector fermion doublet and singlet fermion can mix, and this mixing angle ($\beta$) help us to produce the dark matter density at the right ballpark. We noticed that it is impossible to have {low mass (below $\mathcal{O}(1)$ GeV) range dark matter} without these mixings, either via freeze-in or freeze-out mechanism. Most of the {high-mass (above $\mathcal{O}(1)$ GeV) range dark matter} cannot produce the right relic density and is ruled out from the direct detection experiments. The mixing angle helps us to get the allowed dark matter mass within $keV-TeV$ region.
In this model, we have various channels which can produce dark matter through the decay of the heavy model particles. All the possible decay channels and their partial decay width are given by
\begin{eqnarray}
\Gamma(h\rightarrow \chi_1\chi_1) &=& \frac{M_h}{8 \pi } \, |g_{{\chi_1}\,{\chi_1}\, h }|^2 \left( 1- \frac{4 M_{\chi_1}^2}{M_h^2} \right)^\frac{3}{2},\nn\\
\Gamma(H^\pm \rightarrow \chi_1 l) &=&  \frac{|g_{\overset{-}{l}\,{\chi_1}\,H^+}|^2}{16 \pi M_{H^\pm}^3 } \, \left( M_{H^\pm}^2 - M_{\chi_1}^2 -m_l^2\right) \, \sqrt{  \left( M_{H^\pm}^2 - M_{\chi_1}^2 -m_l^2  \right)^2 - 4 \, M_{\chi_1}^2 \,m_l^2 }, \nn\\
\Gamma(H\rightarrow \chi_1\nu_l) &=&   \frac{|g_{\overset{-}{{\nu_l}}\,{\chi_1}\,{H}}|^2}{16 \pi M_{H}^3 } \, \left( M_{H}^2 - M_{\chi_1}^2\right)^2 , \\
\Gamma(\chi_2 \rightarrow \chi_1 h) &=&   \frac{|g_{{\chi_1}\,{\chi_2}\, h}|^2  }{16 \pi M_{\chi_2}^3} \,  \left( M_{\chi_2}^2 - M_{\chi_1}^2 -M_h^2\right) \, \sqrt{  \left( M_{\chi_2}^2 - M_{\chi_1}^2 -M_h^2  \right)^2 - 4 \, M_{\chi_1}^2 \,M_h^2 }, \nn \\
\Gamma(\chi_2 \rightarrow \chi_1 Z) &=&   \frac{|{ {cb}\, \,{sb}\, \sqrt{g_1^2+g_2^2}}/{2}|^2 }{16 \pi  M_{\chi_2}^3 } \,  \sqrt{  \left( M_{\chi_2}^2 - M_{\chi_1}^2 -M_Z^2  \right)^2 - 4 \, M_{\chi_1}^2 \,M_Z^2 }, \nn\\
&&\times \, 2 \left( M_{\chi_2}^2 + M_{\chi_1}^2 + 2 M_{\chi_2} M_{\chi_1} -M_Z^2 \right),\nn\\
\Gamma(E_1^\pm \rightarrow \chi_1 W) &=&  \frac{|{ g_2 \,{sb} }/{\sqrt{2} }|^2  }{16 \pi M_{\chi_2}^3 } \,  \,  \sqrt{  \left( M_{\chi_2}^2 - M_{\chi_1}^2 -M_W^2  \right)^2 - 4 \, M_{\chi_1}^2 \,M_W^2 }, \nn\\
&&\times \, 2 \left( M_{\chi_2}^2 + M_{\chi_1}^2 + 2 M_{\chi_2} M_{\chi_1} -M_W^2 \right),\nn\\
\Gamma(Z\rightarrow \chi_1\chi_1) &=&  \frac{|{ {sb}\,^2 \sqrt{g_1^2+g_2^2} }/{2 }|^2  \, M_Z}{48 \pi }. \,  \nn
\label{eq:decay}
\end{eqnarray}
The coupling strength are defined as:
\begin{eqnarray}
& g_{{\chi_1}\,{\chi_1}\, h }={2 i \,{cb}\,^2 \,{sb}\,^2 \, \Delta M_{\chi} }/{{v_{SM}}},  \, g_{{\chi_1}\,{\chi_2}\, h}  ={i \,{cb}\, \,{sb}\, \left(\,{cb}\,^2-\,{sb}\,^2\right) \Delta M_{\chi} }/{{v_{SM}}},\nn\\
& g_{\overset{-}{l}\,{\chi_1}\,H^+}=-i \,{sb}\,  Y_{fl1}, \, g_{\overset{-}{{\nu_l}}\,{\chi_1}\,{H}}=-{i \,{sb}\,  Y_{fl1}}/{\sqrt{2}},\, g_{{\chi_1}\,{E1}\, W  }={i g_2 \,{sb}\, \gamma^{\mu}}/{\sqrt{2} },\\
&g_{{\chi_1}\,{\chi_2}\, Z}=-{i \,{cb}\, \,{sb}\, \sqrt{g_1^2+g_2^2} \, \gamma^{\mu}\gamma^5}/{2} ,\, g_{{\chi_1}\,{\chi_1}\, Z}=-{i \,{sb}\,^2 \sqrt{g_1^2+g_2^2} \,\gamma^{\mu}\gamma^5}/{2 }, \nn
\label{eq:coups}
\end{eqnarray}
where, $\Delta M_{\chi}= M_{\chi_2}-M_{\chi_1}$, $cb\equiv\cos\beta$ and $sb\equiv\sin\beta$.

\begin{table*}[h!]
	\centering
	\begin{tabular}{|p{1.6cm}|p{2.5cm}|p{2.5cm}|p{3cm}|c|p{4.7cm}|}
		\hline
		\hline
		Channel & $~~ M_{\chi_1}$ (GeV) & $~~ M_{\chi_2}$ (GeV) & ~~$~~~sin\beta~~~$&$\Omega_{DM}h^2$&~~~~~~~~~~Percentage \\
		\hline

		      &&&&&$\Gamma(\chi_2\rightarrow \chi_1 h)\quad~93.56\%$\\
			~~BP-1&~~$10^{-5}$ & ~~~1000& $~~6.006\times 10^{-9}$&0.1194& $\Gamma(\chi_2\rightarrow \chi_1 Z)\quad 1.58 \%$\\
			 &&&&&$\Gamma({E_1^\pm}\rightarrow \chi_1 W)\quad~4.86\%$\\
		\hline

		      &&&&&$\Gamma(\chi_2\rightarrow \chi_1 h)\quad~93.58\%$\\
			~~BP-2&~~$10^{-3}$ & ~~~1000& $~~6.015\times 10^{-10}$&0.1198& $\Gamma(\chi_2\rightarrow \chi_1 Z)\quad 1.60 \%$\\
			 &&&&&$\Gamma({E_1^\pm}\rightarrow \chi_1 W)\quad~4.82\%$\\
		\hline

		      &&&&&$\Gamma(\chi_2\rightarrow \chi_1 h)\quad~93.58\%$\\
			~~BP-4&~~$5$ & ~~~1000& $~~8.543\times 10^{-12}$&0.1194& $\Gamma(\chi_2\rightarrow \chi_1 Z)\quad 1.60 \%$\\
			 &&&&&$\Gamma({E_1^\pm}\rightarrow \chi_1 W)\quad~4.82\%$\\
		\hline

		      &&&&&$\Gamma(\chi_2\rightarrow \chi_1 h)\quad~86.07\%$\\
			~~BP-5&~~$200$ & ~~~1000& $~~1.626\times 10^{-12}$&0.112& $\Gamma(\chi_2\rightarrow \chi_1 Z)\quad 3.42 \%$\\
			 &&&&&$\Gamma({E_1^\pm}\rightarrow \chi_1 W)\quad~10.51\%$\\
		\hline

		      &&&&&$\Gamma(\chi_2\rightarrow \chi_1 h)\quad~35.94\%$\\
			~~BP-6&~~$600$ & ~~~1000& $~~1.967\times 10^{-12}$&0.127& $\Gamma(\chi_2\rightarrow \chi_1 Z)\quad 15.69 \%$\\
			 &&&&&$\Gamma({E_1^\pm}\rightarrow \chi_1 W)\quad~48.35\%$\\
		\hline

		      &&&&&$\Gamma(\chi_2\rightarrow \chi_1 h)\quad~93.94\%$\\
			~~BP-7&~~$600$ & ~~~2000& $~~8.549\times 10^{-13}$&0.127& $\Gamma(\chi_2\rightarrow \chi_1 Z)\quad 1.49 \%$\\
			 &&&&&$\Gamma({E_1^\pm}\rightarrow \chi_1 W)\quad~4.567\%$\\
		\hline

		      &&&&&$\Gamma(\chi_2\rightarrow \chi_1 h)\quad~33.96\%$\\
			~~BP-8&~~$1500$ & ~~~2000& $~~1.916\times 10^{-12}$&0.119& $\Gamma(\chi_2\rightarrow \chi_1 Z)\quad 16.02 \%$\\
			 &&&&&$\Gamma({E_1^\pm}\rightarrow \chi_1 W)\quad~48.92\%$\\
		\hline

		      &&&&&$\Gamma(\chi_2\rightarrow \chi_1 h)\quad~2.54\%$\\
			~~BP-9&~~$1800$ & ~~~2000& $~~1.916\times 10^{-12}$&0.112& $\Gamma(\chi_2\rightarrow \chi_1 Z)\quad 23.48 \%$\\
			 &&&&&$\Gamma({E_1^\pm}\rightarrow \chi_1 W)\quad~73.98\%$\\
		\hline
	\end{tabular}
	\caption{The benchmark points allowed by all the theoretical and experimental constraints. We use $M_H=500$ GeV and $Y_{fl,i} ~(i=1,2,3)<0.01$, $\lambda_3\sim 10^{-9}$. These choices of parameters are consistent with neutrino mass, mixing angles and other experimental data.}
	\label{tabDM:BP1}
\end{table*}
 
\begin{figure}[h]
	\centering
	\includegraphics[scale=.30]{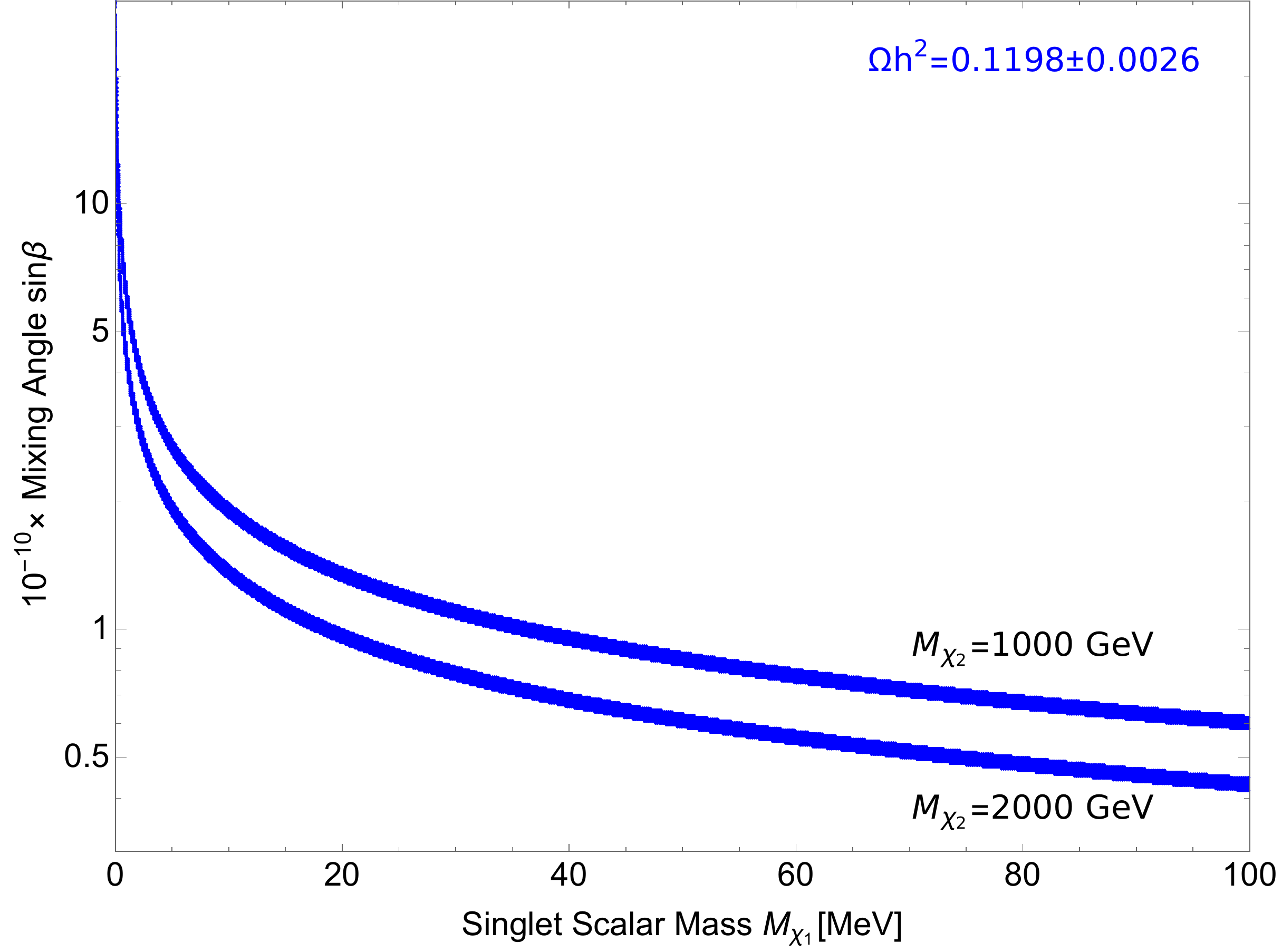}
	\includegraphics[scale=.30]{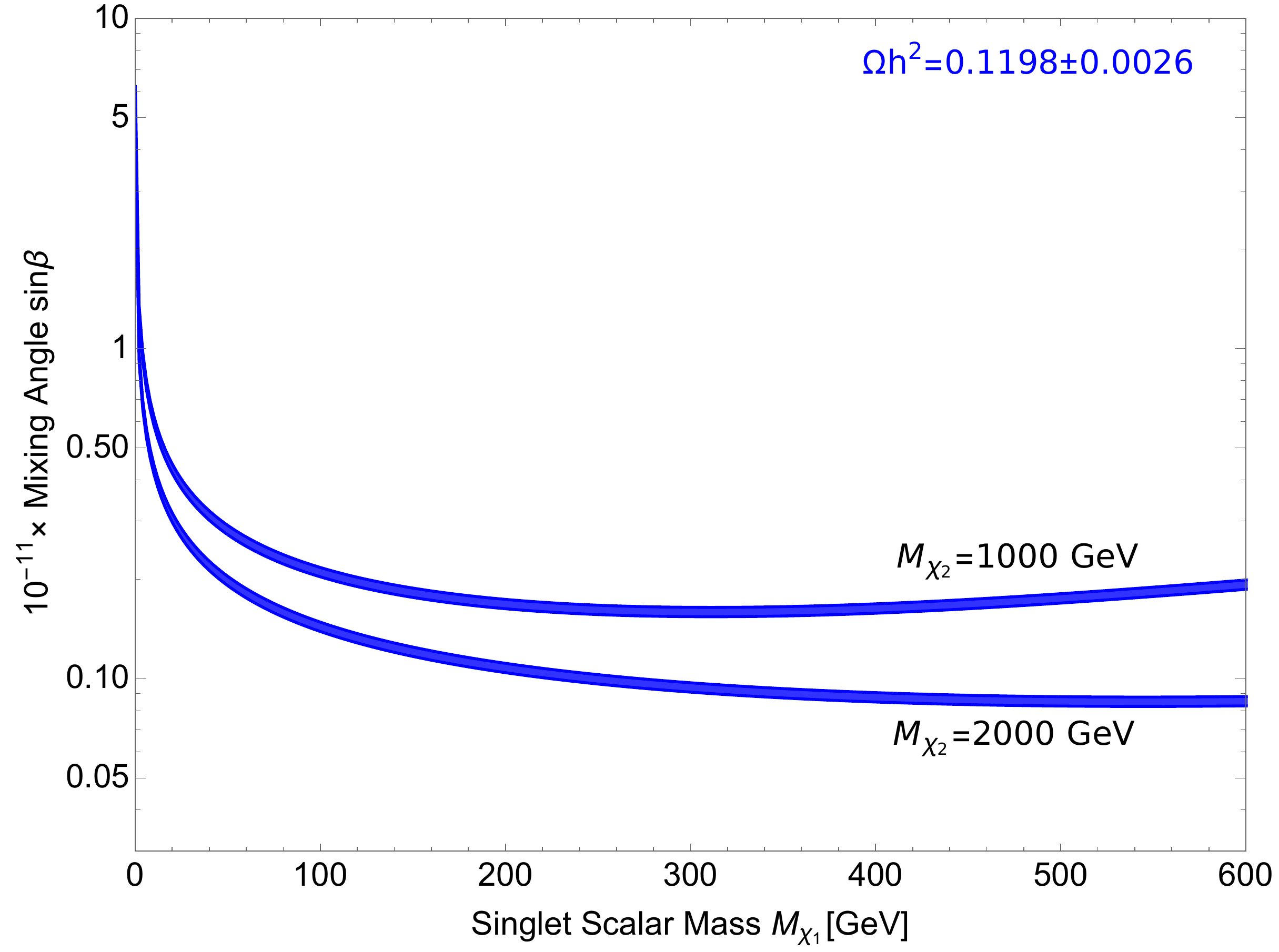}
	\caption{ The blue lines satisfy the exact relic denisty. The upper line stands for the $M_{\chi_2}=1000$ GeV and lower one for  $M_{\chi_2}=2000$ GeV. The slope of these blue lines depend on the mass gap $M_{\chi_2}-M_{\chi_1}$.The triplet Higgs mass $M_{H,H^\pm}$ and Yukawa coupling $Y_{fl1}$ are assumed in such a way that it provide the neutrino mass and mixing angles and allowed by all the theoretical and experimental  constraints (see the table~\ref{tab:neu}).}\label{fig:relicFIMP}
\end{figure}

We show the contour plot~\ref{fig:relicFIMP} in the mixing angle ($\sin\beta$) vs. dark matter mass plane for two different neutral fermion mass $M_{\chi_2}=1000$ GeV and $2000$ GeV respectively. The left plot~\ref{fig:relicFIMP} corresponds to the dark matter mass from $keV-MeV$ range while the $MeV-TeV$ mass range  is shown in the right plot~\ref{fig:relicFIMP}.
We fixed the triplet Higgs mass $M_H$ and Yukawa coupling $Y_{fl1}$ such that it can provide the neutrino low energy variables and allowed by all the theoretical and experimental constraints (e.g., see the table~\ref{tab:neu}).
We find that the $\Gamma(\chi_2 \rightarrow \chi_1 h)$ is always dominant than the other contributions. $\Gamma(h\rightarrow \chi_1\chi_1)$ and $\Gamma(Z\rightarrow \chi_1\chi_1)$ have very tiny contribution. The other channels also have smaller effect as compared to the $\Gamma(\chi_2 \rightarrow \chi_1 h)$. 
The mass difference of the initial heavy and final particles would change the contributions to the relic density.
We show a few related benchmark points, including the percentage of the contributions of various channels in table~\ref{tabDM:BP1} for the neutral fermion mass $M_{\chi_2}=1000$ and $2000$ GeV respectively. We use the heavy Higgs mass at $M_H=500$ GeV and the Yukawa couplings $Y_{fl,i} ~(i=1,2,3)<0.01$ and Higgs portal quartic coupling $\lambda_3\sim 10^{-9}$. Parameters are chosen in such way that consistent with neutrino mass, mixing angles and other experimental data.

It is to be highlighted that the second generation of vector fermion doublets is needed to explain the neutrino mass and mixing angle accurately. Interestingly, these new doublet fermions do not have any mixing; hence the fermions $\chi_3$ and $E_2^\pm$ do not have two-body decay channels with at least one dark matter component in the final state. However, they can decay into dark matter through $>3$-body decay channels, e.g.,  $E_2^\pm \rightarrow H^\pm \nu_l, H^\pm \rightarrow \chi_1 l$ and $\chi_3 \rightarrow H l, H \rightarrow \chi_1 \nu_l$, etc.
We also checked that, these new fermion doublets can annihilate to the dark matter via $Z$ or $h$ bosons, e.g., $\chi_3 \chi_3 \rightarrow Z (h)  \rightarrow \chi_1 \chi_1$ and  $E_2^\pm E_2^\mp \rightarrow Z (h)   \rightarrow \chi_1 \chi_1$, etc. However, the effect of these production from the second doublets are negligibly small as compared to the other production channels. 


\section{Collider signature as charged track}\label{sec:collidr}
{
The tiny fermionic mixing, i.e. $\sin\beta \sim 10^{-11}$ provides the exact relic density for the dark matter candidate in this model.
The direct detection cross-section for the dark matter candidate is much below the XENON-1T bound~\cite{Aprile:2018dbl}.
However, the interactions of the new vector-like fermions with standard model fermions make them more feasible to probe at the high energetic collider. The existing collider searches~\cite{No:2019gvl} includes possible ways to explore dark matter particles. The authors of the Ref.~\cite{No:2019gvl} have primarily investigated the FIMP dark matter candidate in the context of 14 TeV LHC experiments with a future high luminosity at the MATHUSLA surface detector.
In this model, there have a possibility to form a charged track due to the heavy vector like-charged particle $E_1^\pm$ decay into two SM fermions and a dark matter. The decay width of the vector like-charged fermion ($\Gamma(E_1^\pm \rightarrow \chi_1 W^*)$) if proportional to (see eqn.~\ref{eq:decay}) $\sin^2\beta$, we find the decay width $\sim 9\times 10^{-21}$ GeV and corresponding decay length for these charged fermions to be $\mathcal{O}(20)$ Kilometer for this choice of $\sin\beta \sim 10^{-11}$. Now, it is interesting to see if we can get a sufficient number of collider events from these charged tracks for the detection; it mainly depends on the production cross-section of the mother particle $E_1^\pm$ and luminosity. In this model, the dominant production
of the vector fermions come through the Drell-Yan processes. We find the production cross-section of the vector-like charged fermions at $14$ TeV LHC energy for mass $M_{E_1^\pm}=1000$ GeV is $\sim 2$ fb.
The total number of events at the LHC-MATHUSLA can be found in Ref.~\cite{No:2019gvl} and is given by
$
	N_{events} = \sigma^{\rm LHC}_{\sqrt{s}} \, \mathcal{L}  \, \int P^{\rm MATH}_{\rm Decay},~{\rm with}~P^{\rm MATH}_{\rm Decay}=0.05(e^{-\frac{L_a}{\beta \, c \, \tau_{E_1^\pm}}} - e^{-\frac{L_b}{\beta \, c \, \tau_{E_1^\pm}}}).$
Hence it might need larger luminosity and/or energy to get significant events at the present MATHUSLA surface detector for our analysis.
For this choice of $\sin\beta \sim 10^{-11}$, we find $N_{events}>3$ at 14 TeV LHC with an integrated luminosity $\mathcal{L} = 5.0\times 10^4 ~{\rm fb}^{-1}$.}

\section{Discussion and Conclusion}
\label{sec:conclu}
We study the possibility of FIMP dark matter in an extended hyperchargeless Higgs triplet model in this work.
This model contains additional two vector-like doublet fermions and a neutral singlet fermion. 
We also analyze the neutrino sector keeping an eye on the latest muon anomalous magnetic moment data along with dark matter study. A detailed discussion on this model has not yet explored with these recent results, which motivates us to carry out our study.

We impose a discrete $Z_2$ symmetry such that the scalar triplet and additional fermions are $odd$ under this $Z_2$ transformation while SM fields do retain even.
The extra scalar triplet consists of a pair of singly-charged fields $H^\pm$ and a $CP$-even neutral scalar field $H$. The neutral component of the first vector-like fermion doublet can mix with the neutral singlet fermion, and the lightest eigenstate (mainly composed of the neutral singlet fermion) behaves as a viable dark matter candidate, and this mixing angle will help us to get the dark matter density through Freeze-in mechanism.
We assume that the second vector-like fermion does not mix with the other two new fermions; however, it contributes to neutrino mass and mixing angles.
The other fermion mass eigenstates and the neutral component of the triplet scalar $H$ can also serve as dark matter candidate depending on the mass. In such a case, one can get the relic density through the freeze-out mechanism only. However, for the fermionic dark matter candidate (mostly composed of neutral component of first vector doublet) case, $M_{\chi_2}<M_{\chi_{1,3}, E^\pm_{1,2}, H}$, and { most of the regions are ruled out by the present direct detection experiments. It has additional $Z$-portal diagram for the direct searches.}
One can get the scalar dark matter for $M_{H}<M_{\chi_{1,2,3}, E^\pm_{1,2}}$. In an earlier study~\cite{Khan:2016sxm}, it was  found that the exact relic density can be found at $M_{H} \sim 2 $ TeV. The low mass region gives under abundance due to the large co-annihilation channels $H H^\pm \rightarrow W \rightarrow SM~ particles$. We also see that the additional enhancement come from the diagram $H H^\pm \, (H H) \rightarrow SM~ leptons$ through $t$-channels which having $\chi_{1,2,3},E^\pm_{1,2}$ propagators. {Incidentally, one can get relic density $M_{H} \sim 2.8 $ TeV through freeze-out mechanism only.} In such a scenario, the low mass region for the scalar dark matter gives under abundance; conversely, it is ruled out from the relic density and/or direct detection for doublet-type neutral fermion. These components also ruled out as a FIMP dark matter candidate,
give the overabundance of the dark matter relic density through the Freeze-in mechanism. We also check that $\chi_1$ (mostly compose of singlet neutral fermion) as a WIMP dark matter is ruled out as it gives overabundance through   
Freeze-out mechanism.
Hence, $\chi_1$ is the most suitable dark matter candidate in this model via Freeze-in mechanism, which gives relic density for all mass range $keV-TeV$. 
We show two different plots in the mixing angle $vs.$ dark matter mass plane. It is noted that the allowed dark matter region can also provide the exact neutrino mass and mixing angles by adjusting the other parameters in the model, mainly the Higgs portal coupling $\lambda_3$, Yukawa couplings $Y_{fli}$ for a fixed $M_{\chi_2}$. The effect of these coupling adjustments are negligibly small in the relic density.

This model also discusses the numerical insights to neutrino phenomenology, and neutrino masses achieved via one-loop correction scotogenic model.
A tiny Higgs portal coupling, $\lambda_3$ and/or the Yukawa couplings, $Y_{fli}$ with fermion masses $\mathcal{O}(1)$ TeV can give exact mass order $\sum m_\nu < 0.1$ eV.
We first check with only one generation of the vector-like fermion doublet with every possible way to generate all the neutrino low energy variables.
Eventually, we find that minimum of two doublets are essential to address all the light neutrino parameters. Therefore, we add another doublet without affecting dark matter results. This second fermion doublet extension helps us generate all the neutrino variables in both the normal and inverted hierarchy mass pattern.

In the LFV study, one can have additional corrections (positive or negative depending on the parameters) to the muon and electron anomalous magnetic moments at one-loop level.
The model also has lepton flavor violation decay (especially, $\Gamma (\mu \rightarrow e \gamma)$) through one-loop correction. This decay process provides tight constraints on the model parameters that involve the neutrino mass, dark matter density, and anomalous magnetic moments. We find that the allowed parameter by the LFV process cannot explain the recent muon and electron anomalous magnetic moments data. We need very large new Yukawa couplings (non-perturbative) to explain muon anomalous magnetic moments data, which violates LFV and relic density bounds. Therefore, as a conclusive remark on this model, one can explain the neutrino low energy variables, extensive range of dark matter masses using the parameters allowed by the LFV data with tiny couplings and mixing angles.

\vskip 20pt
\noindent{\bf Acknowledgements:}\\
PD and MKD would like to acknowledge the Department of Science and Technology,  Government of India under project number EMR/2017/001436 for financial aids. 
NK would like to thank Prof. Dilip Kumar Ghosh for the support at IACS-KOLKATA.
NK also would like to thank the organizers of the Invisibles21 virtual Workshop of the European Unions Horizon research and innovation program. This project has received funding /support from the European Union’s Horizon 2020 research and innovation programme under the Marie Skłodowska -Curie grant agreement No 860881-HIDDeN”

\appendix

\section{Mixing of three new neutral neutral and charged fermions}
\label{app:A}
If we consider, the effect of the second generation of fermion, then we mast need to diagonalize the following matrix.
\begin{eqnarray}
\mathcal{M}=\begin{pmatrix}
M_{NS}&M_X^1 & M_X^2\\
M_X^{1,\dagger}&M_{ND}^{11} & M_{ND}^{12}\\
M_X^{2,\dagger}& M_{ND}^{21} & M_{ND}^{22}\\
\end{pmatrix},
\label{eqn:mass}
\end{eqnarray}
where, $M_X^1=\frac{v \, Y_{N,1} }{\sqrt{2}}$ and $M_X^2=\frac{v \, Y_{N,2} \,}{\sqrt{2}}$. The neutral component of the fermion doublets ($N_{D,1}$ and $N_{D,2}$) and the singlet charged fermion ($ {N}_S$) mix at tree level. Let us assume the second VLF doublet is not decoupled, i.e., $Y_{N,2}\neq 0$ and $M_{ND}^{21}=M_{ND}^{12}\neq0$. Then,
the mass eigenstates are obtained by diagonalizing the mass matrix (similar to the Pontecorvo-Maki-Nakagawa-Sakata (PMNS) matrix with $zero$ phases) with
a rotation of the  $ ({N}_S$  $N_{D,1}$ $N_{D,2})$  basis as,
\begin{eqnarray}
\begin{pmatrix}
\chi_1 \\\chi_2\\ \chi_3\\
\end{pmatrix}=U_{PMNS}\begin{pmatrix}
N_S\\ N_{D,1} \\ N_{D,2}\\
\end{pmatrix},
\end{eqnarray}
where,
\begin{eqnarray}
U_{PMNS}=\begin{pmatrix}
\cos\beta_{12}&\sin\beta_{12} & 0\\
-\sin\beta_{12}&\cos\beta_{12} &0\\
0 & 0 & 1\\
\end{pmatrix}  \begin{pmatrix}
\cos\beta_{13}&0&\sin\beta_{13} \\
0 & 1 & 0\\
-\sin\beta_{13}&0&\cos\beta_{13}\\
\end{pmatrix}  \begin{pmatrix}
1 & 0 & 0\\
0&\cos\beta_{23}&\sin\beta_{23} \\
0&-\sin\beta_{23}&\cos\beta_{23}\\
\end{pmatrix} .
\end{eqnarray}
The new charged fermion also do get mixed together and the mass matrix can be written as:
\begin{eqnarray}
\mathcal{M}_{c}=\begin{pmatrix}
M_{ND}^{11} & M_{ND}^{12}\\
M_{ND}^{21} & M_{ND}^{22}\\
\end{pmatrix},
\label{eqn:mass2}
\end{eqnarray}
With the condition $M_{ND}^{12}=M_{ND}^{21}$, the mass eigenstates are obtained by diagonalizing the mass matrix with
a rotation of the ( $E_{D,1}^\pm$ $E_{D,2}^\pm$ ) basis as:
\begin{eqnarray}
\begin{pmatrix}
E_1^\pm \\ E_2^\pm\\
\end{pmatrix}=\begin{pmatrix}
\cos\beta_{23}&\sin\beta_{23} \\
-\sin\beta_{23}&\cos\beta_{23}\\
\end{pmatrix} \begin{pmatrix}
E_{D,1}^\pm \\ E_{D,2}^\pm \\
\end{pmatrix}, {~\rm with~} \tan 2 \beta_{23} = \frac{2 M_{ND}^{12}}{M_{ND}^{22}-M_{ND}^{11}}.
\end{eqnarray}
The masses for $M_{ND}^{22}-M_{ND}^{11}>>M_{ND}^{12}$ can be written as:
\begin{eqnarray}
M_{E_1^\pm} &=& M_{ND}^{11} - \frac{2 (M_{ND}^{12})^2}{M_{ND}^{22}-M_{ND}^{11}}, ~{\rm and}~~M_{E_2^\pm} =  M_{ND}^{22} + \frac{2 (M_{ND}^{12})^2}{M_{ND}^{22}-M_{ND}^{11}}.
\end{eqnarray}
All the non-zero of these off-diagonal term could change our results which we left for our future analysis.
Please note that the limits $\beta_{13}=\beta_{23}=0$ produce our results.

\bibliographystyle{apsrev4-1}
\bibliography{tevportalnew}

\begin{thebibliography}{93}%
\makeatletter
\providecommand \@ifxundefined [1]{%
 \@ifx{#1\undefined}
}%
\providecommand \@ifnum [1]{%
 \ifnum #1\expandafter \@firstoftwo
 \else \expandafter \@secondoftwo
 \fi
}%
\providecommand \@ifx [1]{%
 \ifx #1\expandafter \@firstoftwo
 \else \expandafter \@secondoftwo
 \fi
}%
\providecommand \natexlab [1]{#1}%
\providecommand \enquote  [1]{``#1''}%
\providecommand \bibnamefont  [1]{#1}%
\providecommand \bibfnamefont [1]{#1}%
\providecommand \citenamefont [1]{#1}%
\providecommand \href@noop [0]{\@secondoftwo}%
\providecommand \href [0]{\begingroup \@sanitize@url \@href}%
\providecommand \@href[1]{\@@startlink{#1}\@@href}%
\providecommand \@@href[1]{\endgroup#1\@@endlink}%
\providecommand \@sanitize@url [0]{\catcode `\\12\catcode `\$12\catcode
  `\&12\catcode `\#12\catcode `\^12\catcode `\_12\catcode `\%12\relax}%
\providecommand \@@startlink[1]{}%
\providecommand \@@endlink[0]{}%
\providecommand \url  [0]{\begingroup\@sanitize@url \@url }%
\providecommand \@url [1]{\endgroup\@href {#1}{\urlprefix }}%
\providecommand \urlprefix  [0]{URL }%
\providecommand \Eprint [0]{\href }%
\providecommand \doibase [0]{http://dx.doi.org/}%
\providecommand \selectlanguage [0]{\@gobble}%
\providecommand \bibinfo  [0]{\@secondoftwo}%
\providecommand \bibfield  [0]{\@secondoftwo}%
\providecommand \translation [1]{[#1]}%
\providecommand \BibitemOpen [0]{}%
\providecommand \bibitemStop [0]{}%
\providecommand \bibitemNoStop [0]{.\EOS\space}%
\providecommand \EOS [0]{\spacefactor3000\relax}%
\providecommand \BibitemShut  [1]{\csname bibitem#1\endcsname}%
\let\auto@bib@innerbib\@empty
\bibitem [{\citenamefont {Aghanim}\ \emph {et~al.}(2018)\citenamefont {Aghanim}
  \emph {et~al.}}]{Aghanim:2018eyx}%
  \BibitemOpen
  \bibfield  {author} {\bibinfo {author} {\bibfnamefont {N.}~\bibnamefont
  {Aghanim}} \emph {et~al.} (\bibinfo {collaboration} {Planck}),\ }\href@noop
  {} {\  (\bibinfo {year} {2018})},\ \Eprint {http://arxiv.org/abs/1807.06209}
  {arXiv:1807.06209 [astro-ph.CO]} \BibitemShut {NoStop}%
\bibitem [{\citenamefont {Bennett}\ \emph {et~al.}(2013)\citenamefont {Bennett}
  \emph {et~al.}}]{Bennett:2012zja}%
  \BibitemOpen
  \bibfield  {author} {\bibinfo {author} {\bibfnamefont {C.~L.}\ \bibnamefont
  {Bennett}} \emph {et~al.} (\bibinfo {collaboration} {WMAP}),\ }\href
  {\doibase 10.1088/0067-0049/208/2/20} {\bibfield  {journal} {\bibinfo
  {journal} {Astrophys. J. Suppl.}\ }\textbf {\bibinfo {volume} {208}},\
  \bibinfo {pages} {20} (\bibinfo {year} {2013})},\ \Eprint
  {http://arxiv.org/abs/1212.5225} {arXiv:1212.5225 [astro-ph.CO]} \BibitemShut
  {NoStop}%
\bibitem [{\citenamefont {Zwicky}(1937)}]{Zwicky:1937zza}%
  \BibitemOpen
  \bibfield  {author} {\bibinfo {author} {\bibfnamefont {F.}~\bibnamefont
  {Zwicky}},\ }\href {\doibase 10.1086/143864} {\bibfield  {journal} {\bibinfo
  {journal} {Astrophys. J.}\ }\textbf {\bibinfo {volume} {86}},\ \bibinfo
  {pages} {217} (\bibinfo {year} {1937})}\BibitemShut {NoStop}%
\bibitem [{\citenamefont {Freese}(2009)}]{Freese:2008cz}%
  \BibitemOpen
  \bibfield  {author} {\bibinfo {author} {\bibfnamefont {K.}~\bibnamefont
  {Freese}},\ }\bibfield  {booktitle} {\emph {\bibinfo {booktitle}
  {{Proceedings, Dark Energy and Dark Matter: Observations Experiments and
  Theories: Lyon, France, July 7-11, 2008}}},\ }\href {\doibase
  10.1051/eas/0936016} {\bibfield  {journal} {\bibinfo  {journal} {EAS Publ.
  Ser.}\ }\textbf {\bibinfo {volume} {36}},\ \bibinfo {pages} {113} (\bibinfo
  {year} {2009})},\ \Eprint {http://arxiv.org/abs/0812.4005} {arXiv:0812.4005
  [astro-ph]} \BibitemShut {NoStop}%
\bibitem [{\citenamefont {Clowe}\ \emph {et~al.}(2006)\citenamefont {Clowe},
  \citenamefont {Bradac}, \citenamefont {Gonzalez}, \citenamefont {Markevitch},
  \citenamefont {Randall}, \citenamefont {Jones},\ and\ \citenamefont
  {Zaritsky}}]{Clowe:2006eq}%
  \BibitemOpen
  \bibfield  {author} {\bibinfo {author} {\bibfnamefont {D.}~\bibnamefont
  {Clowe}}, \bibinfo {author} {\bibfnamefont {M.}~\bibnamefont {Bradac}},
  \bibinfo {author} {\bibfnamefont {A.~H.}\ \bibnamefont {Gonzalez}}, \bibinfo
  {author} {\bibfnamefont {M.}~\bibnamefont {Markevitch}}, \bibinfo {author}
  {\bibfnamefont {S.~W.}\ \bibnamefont {Randall}}, \bibinfo {author}
  {\bibfnamefont {C.}~\bibnamefont {Jones}}, \ and\ \bibinfo {author}
  {\bibfnamefont {D.}~\bibnamefont {Zaritsky}},\ }\href {\doibase
  10.1086/508162} {\bibfield  {journal} {\bibinfo  {journal} {Astrophys. J.
  Lett.}\ }\textbf {\bibinfo {volume} {648}},\ \bibinfo {pages} {L109}
  (\bibinfo {year} {2006})},\ \Eprint {http://arxiv.org/abs/astro-ph/0608407}
  {arXiv:astro-ph/0608407} \BibitemShut {NoStop}%
\bibitem [{\citenamefont {Kolb}\ and\ \citenamefont
  {Turner}(1990)}]{Kolb:1990vq}%
  \BibitemOpen
  \bibfield  {author} {\bibinfo {author} {\bibfnamefont {E.~W.}\ \bibnamefont
  {Kolb}}\ and\ \bibinfo {author} {\bibfnamefont {M.~S.}\ \bibnamefont
  {Turner}},\ }\href@noop {} {\emph {\bibinfo {title} {{The Early
  Universe}}}},\ Vol.~\bibinfo {volume} {69}\ (\bibinfo {year}
  {1990})\BibitemShut {NoStop}%
\bibitem [{\citenamefont {Hall}\ \emph {et~al.}(2010)\citenamefont {Hall},
  \citenamefont {Jedamzik}, \citenamefont {March-Russell},\ and\ \citenamefont
  {West}}]{Hall:2009bx}%
  \BibitemOpen
  \bibfield  {author} {\bibinfo {author} {\bibfnamefont {L.~J.}\ \bibnamefont
  {Hall}}, \bibinfo {author} {\bibfnamefont {K.}~\bibnamefont {Jedamzik}},
  \bibinfo {author} {\bibfnamefont {J.}~\bibnamefont {March-Russell}}, \ and\
  \bibinfo {author} {\bibfnamefont {S.~M.}\ \bibnamefont {West}},\ }\href
  {\doibase 10.1007/JHEP03(2010)080} {\bibfield  {journal} {\bibinfo  {journal}
  {JHEP}\ }\textbf {\bibinfo {volume} {03}},\ \bibinfo {pages} {080} (\bibinfo
  {year} {2010})},\ \Eprint {http://arxiv.org/abs/0911.1120} {arXiv:0911.1120
  [hep-ph]} \BibitemShut {NoStop}%
\bibitem [{\citenamefont {Murayama}(2007)}]{Murayama:2007ek}%
  \BibitemOpen
  \bibfield  {author} {\bibinfo {author} {\bibfnamefont {H.}~\bibnamefont
  {Murayama}},\ }in\ \href@noop {} {\emph {\bibinfo {booktitle} {{Les Houches
  Summer School - Session 86: Particle Physics and Cosmology: The Fabric of
  Spacetime}}}}\ (\bibinfo {year} {2007})\ \Eprint
  {http://arxiv.org/abs/0704.2276} {arXiv:0704.2276 [hep-ph]} \BibitemShut
  {NoStop}%
\bibitem [{\citenamefont {Abe}\ \emph {et~al.}(2016)\citenamefont {Abe} \emph
  {et~al.}}]{Abe:2016nxk}%
  \BibitemOpen
  \bibfield  {author} {\bibinfo {author} {\bibfnamefont {K.}~\bibnamefont
  {Abe}} \emph {et~al.} (\bibinfo {collaboration} {Super-Kamiokande}),\ }\href
  {\doibase 10.1103/PhysRevD.94.052010} {\bibfield  {journal} {\bibinfo
  {journal} {Phys. Rev.}\ }\textbf {\bibinfo {volume} {D94}},\ \bibinfo {pages}
  {052010} (\bibinfo {year} {2016})},\ \Eprint
  {http://arxiv.org/abs/1606.07538} {arXiv:1606.07538 [hep-ex]} \BibitemShut
  {NoStop}%
\bibitem [{\citenamefont {An}\ \emph {et~al.}(2012)\citenamefont {An} \emph
  {et~al.}}]{An:2012eh}%
  \BibitemOpen
  \bibfield  {author} {\bibinfo {author} {\bibfnamefont {F.~P.}\ \bibnamefont
  {An}} \emph {et~al.} (\bibinfo {collaboration} {Daya Bay}),\ }\href {\doibase
  10.1103/PhysRevLett.108.171803} {\bibfield  {journal} {\bibinfo  {journal}
  {Phys. Rev. Lett.}\ }\textbf {\bibinfo {volume} {108}},\ \bibinfo {pages}
  {171803} (\bibinfo {year} {2012})},\ \Eprint {http://arxiv.org/abs/1203.1669}
  {arXiv:1203.1669 [hep-ex]} \BibitemShut {NoStop}%
\bibitem [{\citenamefont {Abe}\ \emph {et~al.}(2012)\citenamefont {Abe} \emph
  {et~al.}}]{Abe:2011fz}%
  \BibitemOpen
  \bibfield  {author} {\bibinfo {author} {\bibfnamefont {Y.}~\bibnamefont
  {Abe}} \emph {et~al.} (\bibinfo {collaboration} {Double Chooz}),\ }\href
  {\doibase 10.1103/PhysRevLett.108.131801} {\bibfield  {journal} {\bibinfo
  {journal} {Phys. Rev. Lett.}\ }\textbf {\bibinfo {volume} {108}},\ \bibinfo
  {pages} {131801} (\bibinfo {year} {2012})},\ \Eprint
  {http://arxiv.org/abs/1112.6353} {arXiv:1112.6353 [hep-ex]} \BibitemShut
  {NoStop}%
\bibitem [{\citenamefont {Roy~Choudhury}\ and\ \citenamefont
  {Choubey}(2018)}]{Choudhury:2018byy}%
  \BibitemOpen
  \bibfield  {author} {\bibinfo {author} {\bibfnamefont {S.}~\bibnamefont
  {Roy~Choudhury}}\ and\ \bibinfo {author} {\bibfnamefont {S.}~\bibnamefont
  {Choubey}},\ }\href {\doibase 10.1088/1475-7516/2018/09/017} {\bibfield
  {journal} {\bibinfo  {journal} {JCAP}\ }\textbf {\bibinfo {volume} {1809}},\
  \bibinfo {pages} {017} (\bibinfo {year} {2018})},\ \Eprint
  {http://arxiv.org/abs/1806.10832} {arXiv:1806.10832 [astro-ph.CO]}
  \BibitemShut {NoStop}%
\bibitem [{\citenamefont {Sirunyan}\ \emph {et~al.}(2018)\citenamefont
  {Sirunyan} \emph {et~al.}}]{Sirunyan:2017khh}%
  \BibitemOpen
  \bibfield  {author} {\bibinfo {author} {\bibfnamefont {A.~M.}\ \bibnamefont
  {Sirunyan}} \emph {et~al.} (\bibinfo {collaboration} {CMS}),\ }\href
  {\doibase 10.1016/j.physletb.2018.02.004} {\bibfield  {journal} {\bibinfo
  {journal} {Phys. Lett.}\ }\textbf {\bibinfo {volume} {B779}},\ \bibinfo
  {pages} {283} (\bibinfo {year} {2018})},\ \Eprint
  {http://arxiv.org/abs/1708.00373} {arXiv:1708.00373 [hep-ex]} \BibitemShut
  {NoStop}%
\bibitem [{\citenamefont {Sirunyan}\ \emph
  {et~al.}(2019{\natexlab{a}})\citenamefont {Sirunyan} \emph
  {et~al.}}]{Sirunyan:2018koj}%
  \BibitemOpen
  \bibfield  {author} {\bibinfo {author} {\bibfnamefont {A.~M.}\ \bibnamefont
  {Sirunyan}} \emph {et~al.} (\bibinfo {collaboration} {CMS}),\ }\href
  {\doibase 10.1140/epjc/s10052-019-6909-y} {\bibfield  {journal} {\bibinfo
  {journal} {Eur. Phys. J.}\ }\textbf {\bibinfo {volume} {C79}},\ \bibinfo
  {pages} {421} (\bibinfo {year} {2019}{\natexlab{a}})},\ \Eprint
  {http://arxiv.org/abs/1809.10733} {arXiv:1809.10733 [hep-ex]} \BibitemShut
  {NoStop}%
\bibitem [{\citenamefont {Mohapatra}\ and\ \citenamefont
  {Senjanovic}(1980)}]{Mohapatra:1979ia}%
  \BibitemOpen
  \bibfield  {author} {\bibinfo {author} {\bibfnamefont {R.~N.}\ \bibnamefont
  {Mohapatra}}\ and\ \bibinfo {author} {\bibfnamefont {G.}~\bibnamefont
  {Senjanovic}},\ }\href {\doibase 10.1103/PhysRevLett.44.912} {\bibfield
  {journal} {\bibinfo  {journal} {Phys. Rev. Lett.}\ }\textbf {\bibinfo
  {volume} {44}},\ \bibinfo {pages} {912} (\bibinfo {year} {1980})},\ \bibinfo
  {note} {[,231(1979)]}\BibitemShut {NoStop}%
\bibitem [{\citenamefont {Bernabeu}\ \emph {et~al.}(1987)\citenamefont
  {Bernabeu}, \citenamefont {Santamaria}, \citenamefont {Vidal}, \citenamefont
  {Mendez},\ and\ \citenamefont {Valle}}]{Bernabeu:1987gr}%
  \BibitemOpen
  \bibfield  {author} {\bibinfo {author} {\bibfnamefont {J.}~\bibnamefont
  {Bernabeu}}, \bibinfo {author} {\bibfnamefont {A.}~\bibnamefont
  {Santamaria}}, \bibinfo {author} {\bibfnamefont {J.}~\bibnamefont {Vidal}},
  \bibinfo {author} {\bibfnamefont {A.}~\bibnamefont {Mendez}}, \ and\ \bibinfo
  {author} {\bibfnamefont {J.~W.~F.}\ \bibnamefont {Valle}},\ }\href {\doibase
  10.1016/0370-2693(87)91100-2} {\bibfield  {journal} {\bibinfo  {journal}
  {Phys. Lett. B}\ }\textbf {\bibinfo {volume} {187}},\ \bibinfo {pages} {303}
  (\bibinfo {year} {1987})}\BibitemShut {NoStop}%
\bibitem [{\citenamefont {Babu}(2010)}]{Babu:2009fd}%
  \BibitemOpen
  \bibfield  {author} {\bibinfo {author} {\bibfnamefont {K.~S.}\ \bibnamefont
  {Babu}},\ }in\ \href {\doibase 10.1142/9789812838360_0002} {\emph {\bibinfo
  {booktitle} {{Proceedings of Theoretical Advanced Study Institute in
  Elementary Particle Physics on The dawn of the LHC era (TASI 2008): Boulder,
  USA, June 2-27, 2008}}}}\ (\bibinfo {year} {2010})\ pp.\ \bibinfo {pages}
  {49--123},\ \Eprint {http://arxiv.org/abs/0910.2948} {arXiv:0910.2948
  [hep-ph]} \BibitemShut {NoStop}%
\bibitem [{\citenamefont {Borah}\ \emph {et~al.}(2018)\citenamefont {Borah},
  \citenamefont {Karmakar},\ and\ \citenamefont {Nanda}}]{Borah:2018gjk}%
  \BibitemOpen
  \bibfield  {author} {\bibinfo {author} {\bibfnamefont {D.}~\bibnamefont
  {Borah}}, \bibinfo {author} {\bibfnamefont {B.}~\bibnamefont {Karmakar}}, \
  and\ \bibinfo {author} {\bibfnamefont {D.}~\bibnamefont {Nanda}},\ }\href
  {\doibase 10.1088/1475-7516/2018/07/039} {\bibfield  {journal} {\bibinfo
  {journal} {JCAP}\ }\textbf {\bibinfo {volume} {07}},\ \bibinfo {pages} {039}
  (\bibinfo {year} {2018})},\ \Eprint {http://arxiv.org/abs/1805.11115}
  {arXiv:1805.11115 [hep-ph]} \BibitemShut {NoStop}%
\bibitem [{\citenamefont {Khan}\ \emph {et~al.}(2014)\citenamefont {Khan},
  \citenamefont {Goswami},\ and\ \citenamefont {Roy}}]{Khan:2012zw}%
  \BibitemOpen
  \bibfield  {author} {\bibinfo {author} {\bibfnamefont {S.}~\bibnamefont
  {Khan}}, \bibinfo {author} {\bibfnamefont {S.}~\bibnamefont {Goswami}}, \
  and\ \bibinfo {author} {\bibfnamefont {S.}~\bibnamefont {Roy}},\ }\href
  {\doibase 10.1103/PhysRevD.89.073021} {\bibfield  {journal} {\bibinfo
  {journal} {Phys. Rev.}\ }\textbf {\bibinfo {volume} {D89}},\ \bibinfo {pages}
  {073021} (\bibinfo {year} {2014})},\ \Eprint {http://arxiv.org/abs/1212.3694}
  {arXiv:1212.3694 [hep-ph]} \BibitemShut {NoStop}%
\bibitem [{\citenamefont {Das}\ \emph {et~al.}(2020)\citenamefont {Das},
  \citenamefont {Das},\ and\ \citenamefont {Khan}}]{Das:2019ntw}%
  \BibitemOpen
  \bibfield  {author} {\bibinfo {author} {\bibfnamefont {P.}~\bibnamefont
  {Das}}, \bibinfo {author} {\bibfnamefont {M.~K.}\ \bibnamefont {Das}}, \ and\
  \bibinfo {author} {\bibfnamefont {N.}~\bibnamefont {Khan}},\ }\href {\doibase
  10.1007/JHEP03(2020)018} {\bibfield  {journal} {\bibinfo  {journal} {JHEP}\
  }\textbf {\bibinfo {volume} {03}},\ \bibinfo {pages} {018} (\bibinfo {year}
  {2020})},\ \Eprint {http://arxiv.org/abs/1911.07243} {arXiv:1911.07243
  [hep-ph]} \BibitemShut {NoStop}%
\bibitem [{\citenamefont {Bhattacharya}\ \emph {et~al.}(2017)\citenamefont
  {Bhattacharya}, \citenamefont {Ghosh}, \citenamefont {Maity},\ and\
  \citenamefont {Ray}}]{Bhattacharya:2017fid}%
  \BibitemOpen
  \bibfield  {author} {\bibinfo {author} {\bibfnamefont {S.}~\bibnamefont
  {Bhattacharya}}, \bibinfo {author} {\bibfnamefont {P.}~\bibnamefont {Ghosh}},
  \bibinfo {author} {\bibfnamefont {T.~N.}\ \bibnamefont {Maity}}, \ and\
  \bibinfo {author} {\bibfnamefont {T.~S.}\ \bibnamefont {Ray}},\ }\href
  {\doibase 10.1007/JHEP10(2017)088} {\bibfield  {journal} {\bibinfo  {journal}
  {JHEP}\ }\textbf {\bibinfo {volume} {10}},\ \bibinfo {pages} {088} (\bibinfo
  {year} {2017})},\ \Eprint {http://arxiv.org/abs/1706.04699} {arXiv:1706.04699
  [hep-ph]} \BibitemShut {NoStop}%
\bibitem [{\citenamefont {Ghosh}\ \emph {et~al.}(2018)\citenamefont {Ghosh},
  \citenamefont {Saha},\ and\ \citenamefont {Sil}}]{Ghosh:2017fmr}%
  \BibitemOpen
  \bibfield  {author} {\bibinfo {author} {\bibfnamefont {P.}~\bibnamefont
  {Ghosh}}, \bibinfo {author} {\bibfnamefont {A.~K.}\ \bibnamefont {Saha}}, \
  and\ \bibinfo {author} {\bibfnamefont {A.}~\bibnamefont {Sil}},\ }\href
  {\doibase 10.1103/PhysRevD.97.075034} {\bibfield  {journal} {\bibinfo
  {journal} {Phys. Rev. D}\ }\textbf {\bibinfo {volume} {97}},\ \bibinfo
  {pages} {075034} (\bibinfo {year} {2018})},\ \Eprint
  {http://arxiv.org/abs/1706.04931} {arXiv:1706.04931 [hep-ph]} \BibitemShut
  {NoStop}%
\bibitem [{\citenamefont {Das}\ \emph {et~al.}(2021{\natexlab{a}})\citenamefont
  {Das}, \citenamefont {Das},\ and\ \citenamefont {Khan}}]{Das:2020hpd}%
  \BibitemOpen
  \bibfield  {author} {\bibinfo {author} {\bibfnamefont {P.}~\bibnamefont
  {Das}}, \bibinfo {author} {\bibfnamefont {M.~K.}\ \bibnamefont {Das}}, \ and\
  \bibinfo {author} {\bibfnamefont {N.}~\bibnamefont {Khan}},\ }\href {\doibase
  10.1016/j.nuclphysb.2021.115307} {\bibfield  {journal} {\bibinfo  {journal}
  {Nucl. Phys. B}\ }\textbf {\bibinfo {volume} {964}},\ \bibinfo {pages}
  {115307} (\bibinfo {year} {2021}{\natexlab{a}})},\ \Eprint
  {http://arxiv.org/abs/2001.04070} {arXiv:2001.04070 [hep-ph]} \BibitemShut
  {NoStop}%
\bibitem [{\citenamefont {Das}(2021)}]{Das:2021xat}%
  \BibitemOpen
  \bibfield  {author} {\bibinfo {author} {\bibfnamefont {P.}~\bibnamefont
  {Das}},\ }\emph {\bibinfo {title} {{Theoretical and phenomenological
  consequences of active and sterile neutrino within Beyond Standard Model
  framework}}},\ \href@noop {} {\bibinfo {type} {Other thesis}} (\bibinfo
  {year} {2021}),\ \Eprint {http://arxiv.org/abs/2107.10622} {arXiv:2107.10622
  [hep-ph]} \BibitemShut {NoStop}%
\bibitem [{\citenamefont {Das}\ and\ \citenamefont {Dey}(2014)}]{Das:2014fea}%
  \BibitemOpen
  \bibfield  {author} {\bibinfo {author} {\bibfnamefont {D.}~\bibnamefont
  {Das}}\ and\ \bibinfo {author} {\bibfnamefont {U.~K.}\ \bibnamefont {Dey}},\
  }\href {\doibase 10.1103/PhysRevD.91.039905, 10.1103/PhysRevD.89.095025}
  {\bibfield  {journal} {\bibinfo  {journal} {Phys. Rev.}\ }\textbf {\bibinfo
  {volume} {D89}},\ \bibinfo {pages} {095025} (\bibinfo {year} {2014})},\
  \bibinfo {note} {[Erratum: Phys. Rev.D91,no.3,039905(2015)]},\ \Eprint
  {http://arxiv.org/abs/1404.2491} {arXiv:1404.2491 [hep-ph]} \BibitemShut
  {NoStop}%
\bibitem [{\citenamefont {Barry}\ \emph {et~al.}(2011)\citenamefont {Barry},
  \citenamefont {Rodejohann},\ and\ \citenamefont {Zhang}}]{Barry:2011wb}%
  \BibitemOpen
  \bibfield  {author} {\bibinfo {author} {\bibfnamefont {J.}~\bibnamefont
  {Barry}}, \bibinfo {author} {\bibfnamefont {W.}~\bibnamefont {Rodejohann}}, \
  and\ \bibinfo {author} {\bibfnamefont {H.}~\bibnamefont {Zhang}},\ }\href
  {\doibase 10.1007/JHEP07(2011)091} {\bibfield  {journal} {\bibinfo  {journal}
  {JHEP}\ }\textbf {\bibinfo {volume} {07}},\ \bibinfo {pages} {091} (\bibinfo
  {year} {2011})},\ \Eprint {http://arxiv.org/abs/1105.3911} {arXiv:1105.3911
  [hep-ph]} \BibitemShut {NoStop}%
\bibitem [{\citenamefont {Burgess}\ \emph {et~al.}(2001)\citenamefont
  {Burgess}, \citenamefont {Pospelov},\ and\ \citenamefont {ter
  Veldhuis}}]{Burgess:2000yq}%
  \BibitemOpen
  \bibfield  {author} {\bibinfo {author} {\bibfnamefont {C.~P.}\ \bibnamefont
  {Burgess}}, \bibinfo {author} {\bibfnamefont {M.}~\bibnamefont {Pospelov}}, \
  and\ \bibinfo {author} {\bibfnamefont {T.}~\bibnamefont {ter Veldhuis}},\
  }\href {\doibase 10.1016/S0550-3213(01)00513-2} {\bibfield  {journal}
  {\bibinfo  {journal} {Nucl. Phys.}\ }\textbf {\bibinfo {volume} {B619}},\
  \bibinfo {pages} {709} (\bibinfo {year} {2001})},\ \Eprint
  {http://arxiv.org/abs/hep-ph/0011335} {arXiv:hep-ph/0011335 [hep-ph]}
  \BibitemShut {NoStop}%
\bibitem [{\citenamefont {Cohen}\ \emph {et~al.}(2012)\citenamefont {Cohen},
  \citenamefont {Kearney}, \citenamefont {Pierce},\ and\ \citenamefont
  {Tucker-Smith}}]{Cohen:2011ec}%
  \BibitemOpen
  \bibfield  {author} {\bibinfo {author} {\bibfnamefont {T.}~\bibnamefont
  {Cohen}}, \bibinfo {author} {\bibfnamefont {J.}~\bibnamefont {Kearney}},
  \bibinfo {author} {\bibfnamefont {A.}~\bibnamefont {Pierce}}, \ and\ \bibinfo
  {author} {\bibfnamefont {D.}~\bibnamefont {Tucker-Smith}},\ }\href {\doibase
  10.1103/PhysRevD.85.075003} {\bibfield  {journal} {\bibinfo  {journal} {Phys.
  Rev.}\ }\textbf {\bibinfo {volume} {D85}},\ \bibinfo {pages} {075003}
  (\bibinfo {year} {2012})},\ \Eprint {http://arxiv.org/abs/1109.2604}
  {arXiv:1109.2604 [hep-ph]} \BibitemShut {NoStop}%
\bibitem [{\citenamefont {Grimus}\ \emph {et~al.}(2009)\citenamefont {Grimus},
  \citenamefont {Lavoura},\ and\ \citenamefont {Radovcic}}]{Grimus:2009mm}%
  \BibitemOpen
  \bibfield  {author} {\bibinfo {author} {\bibfnamefont {W.}~\bibnamefont
  {Grimus}}, \bibinfo {author} {\bibfnamefont {L.}~\bibnamefont {Lavoura}}, \
  and\ \bibinfo {author} {\bibfnamefont {B.}~\bibnamefont {Radovcic}},\ }\href
  {\doibase 10.1016/j.physletb.2009.03.016} {\bibfield  {journal} {\bibinfo
  {journal} {Phys. Lett.}\ }\textbf {\bibinfo {volume} {B674}},\ \bibinfo
  {pages} {117} (\bibinfo {year} {2009})},\ \Eprint
  {http://arxiv.org/abs/0902.2325} {arXiv:0902.2325 [hep-ph]} \BibitemShut
  {NoStop}%
\bibitem [{\citenamefont {Bhupal~Dev}\ \emph {et~al.}(2014)\citenamefont
  {Bhupal~Dev}, \citenamefont {Mazumdar},\ and\ \citenamefont
  {Qutub}}]{Dev:2013yza}%
  \BibitemOpen
  \bibfield  {author} {\bibinfo {author} {\bibfnamefont {P.~S.}\ \bibnamefont
  {Bhupal~Dev}}, \bibinfo {author} {\bibfnamefont {A.}~\bibnamefont
  {Mazumdar}}, \ and\ \bibinfo {author} {\bibfnamefont {S.}~\bibnamefont
  {Qutub}},\ }\href {\doibase 10.3389/fphy.2014.00026} {\bibfield  {journal}
  {\bibinfo  {journal} {Front. in Phys.}\ }\textbf {\bibinfo {volume} {2}},\
  \bibinfo {pages} {26} (\bibinfo {year} {2014})},\ \Eprint
  {http://arxiv.org/abs/1311.5297} {arXiv:1311.5297 [hep-ph]} \BibitemShut
  {NoStop}%
\bibitem [{\citenamefont {Toma}\ and\ \citenamefont
  {Vicente}(2014)}]{Toma:2013zsa}%
  \BibitemOpen
  \bibfield  {author} {\bibinfo {author} {\bibfnamefont {T.}~\bibnamefont
  {Toma}}\ and\ \bibinfo {author} {\bibfnamefont {A.}~\bibnamefont {Vicente}},\
  }\href {\doibase 10.1007/JHEP01(2014)160} {\bibfield  {journal} {\bibinfo
  {journal} {JHEP}\ }\textbf {\bibinfo {volume} {01}},\ \bibinfo {pages} {160}
  (\bibinfo {year} {2014})},\ \Eprint {http://arxiv.org/abs/1312.2840}
  {arXiv:1312.2840 [hep-ph]} \BibitemShut {NoStop}%
\bibitem [{\citenamefont {Ma}(2006)}]{Ma:2006km}%
  \BibitemOpen
  \bibfield  {author} {\bibinfo {author} {\bibfnamefont {E.}~\bibnamefont
  {Ma}},\ }\href {\doibase 10.1103/PhysRevD.73.077301} {\bibfield  {journal}
  {\bibinfo  {journal} {Phys. Rev.}\ }\textbf {\bibinfo {volume} {D73}},\
  \bibinfo {pages} {077301} (\bibinfo {year} {2006})},\ \Eprint
  {http://arxiv.org/abs/hep-ph/0601225} {arXiv:hep-ph/0601225 [hep-ph]}
  \BibitemShut {NoStop}%
\bibitem [{\citenamefont {Weinberg}(1979)}]{weinberg}%
  \BibitemOpen
  \bibfield  {author} {\bibinfo {author} {\bibfnamefont {S.}~\bibnamefont
  {Weinberg}},\ }\href {\doibase 10.1103/PhysRevLett.43.1566} {\bibfield
  {journal} {\bibinfo  {journal} {Phys. Rev. Lett.}\ }\textbf {\bibinfo
  {volume} {43}},\ \bibinfo {pages} {1566} (\bibinfo {year}
  {1979})}\BibitemShut {NoStop}%
\bibitem [{\citenamefont {Gonzalez~Felipe}\ \emph {et~al.}(2004)\citenamefont
  {Gonzalez~Felipe}, \citenamefont {Joaquim},\ and\ \citenamefont
  {Nobre}}]{GonzalezFelipe:2003fi}%
  \BibitemOpen
  \bibfield  {author} {\bibinfo {author} {\bibfnamefont {R.}~\bibnamefont
  {Gonzalez~Felipe}}, \bibinfo {author} {\bibfnamefont {F.~R.}\ \bibnamefont
  {Joaquim}}, \ and\ \bibinfo {author} {\bibfnamefont {B.~M.}\ \bibnamefont
  {Nobre}},\ }\href {\doibase 10.1103/PhysRevD.70.085009} {\bibfield  {journal}
  {\bibinfo  {journal} {Phys. Rev.}\ }\textbf {\bibinfo {volume} {D70}},\
  \bibinfo {pages} {085009} (\bibinfo {year} {2004})},\ \Eprint
  {http://arxiv.org/abs/hep-ph/0311029} {arXiv:hep-ph/0311029 [hep-ph]}
  \BibitemShut {NoStop}%
\bibitem [{\citenamefont {Fraser}\ \emph {et~al.}(2014)\citenamefont {Fraser},
  \citenamefont {Ma},\ and\ \citenamefont {Popov}}]{Fraser:2014yha}%
  \BibitemOpen
  \bibfield  {author} {\bibinfo {author} {\bibfnamefont {S.}~\bibnamefont
  {Fraser}}, \bibinfo {author} {\bibfnamefont {E.}~\bibnamefont {Ma}}, \ and\
  \bibinfo {author} {\bibfnamefont {O.}~\bibnamefont {Popov}},\ }\href
  {\doibase 10.1016/j.physletb.2014.08.069} {\bibfield  {journal} {\bibinfo
  {journal} {Phys. Lett.}\ }\textbf {\bibinfo {volume} {B737}},\ \bibinfo
  {pages} {280} (\bibinfo {year} {2014})},\ \Eprint
  {http://arxiv.org/abs/1408.4785} {arXiv:1408.4785 [hep-ph]} \BibitemShut
  {NoStop}%
\bibitem [{\citenamefont {Merle}\ and\ \citenamefont
  {Platscher}(2015)}]{Merle:2015ica}%
  \BibitemOpen
  \bibfield  {author} {\bibinfo {author} {\bibfnamefont {A.}~\bibnamefont
  {Merle}}\ and\ \bibinfo {author} {\bibfnamefont {M.}~\bibnamefont
  {Platscher}},\ }\href {\doibase 10.1007/JHEP11(2015)148} {\bibfield
  {journal} {\bibinfo  {journal} {JHEP}\ }\textbf {\bibinfo {volume} {11}},\
  \bibinfo {pages} {148} (\bibinfo {year} {2015})},\ \Eprint
  {http://arxiv.org/abs/1507.06314} {arXiv:1507.06314 [hep-ph]} \BibitemShut
  {NoStop}%
\bibitem [{\citenamefont {Law}\ and\ \citenamefont
  {McDonald}(2013)}]{Law:2013saa}%
  \BibitemOpen
  \bibfield  {author} {\bibinfo {author} {\bibfnamefont {S.~S.~C.}\
  \bibnamefont {Law}}\ and\ \bibinfo {author} {\bibfnamefont {K.~L.}\
  \bibnamefont {McDonald}},\ }\href {\doibase 10.1007/JHEP09(2013)092}
  {\bibfield  {journal} {\bibinfo  {journal} {JHEP}\ }\textbf {\bibinfo
  {volume} {09}},\ \bibinfo {pages} {092} (\bibinfo {year} {2013})},\ \Eprint
  {http://arxiv.org/abs/1305.6467} {arXiv:1305.6467 [hep-ph]} \BibitemShut
  {NoStop}%
\bibitem [{\citenamefont {Mahanta}\ and\ \citenamefont
  {Borah}(2019)}]{Mahanta:2019gfe}%
  \BibitemOpen
  \bibfield  {author} {\bibinfo {author} {\bibfnamefont {D.}~\bibnamefont
  {Mahanta}}\ and\ \bibinfo {author} {\bibfnamefont {D.}~\bibnamefont
  {Borah}},\ }\href {\doibase 10.1088/1475-7516/2019/11/021} {\bibfield
  {journal} {\bibinfo  {journal} {JCAP}\ }\textbf {\bibinfo {volume} {1911}},\
  \bibinfo {pages} {021} (\bibinfo {year} {2019})},\ \Eprint
  {http://arxiv.org/abs/1906.03577} {arXiv:1906.03577 [hep-ph]} \BibitemShut
  {NoStop}%
\bibitem [{\citenamefont {Klein}\ \emph {et~al.}(2019)\citenamefont {Klein},
  \citenamefont {Lindner},\ and\ \citenamefont {Ohmer}}]{Klein:2019iws}%
  \BibitemOpen
  \bibfield  {author} {\bibinfo {author} {\bibfnamefont {C.}~\bibnamefont
  {Klein}}, \bibinfo {author} {\bibfnamefont {M.}~\bibnamefont {Lindner}}, \
  and\ \bibinfo {author} {\bibfnamefont {S.}~\bibnamefont {Ohmer}},\ }\href
  {\doibase 10.1007/JHEP03(2019)018} {\bibfield  {journal} {\bibinfo  {journal}
  {JHEP}\ }\textbf {\bibinfo {volume} {03}},\ \bibinfo {pages} {018} (\bibinfo
  {year} {2019})},\ \Eprint {http://arxiv.org/abs/1901.03225} {arXiv:1901.03225
  [hep-ph]} \BibitemShut {NoStop}%
\bibitem [{\citenamefont {Sarma}\ \emph {et~al.}(2021)\citenamefont {Sarma},
  \citenamefont {Das},\ and\ \citenamefont {Das}}]{Sarma:2020msa}%
  \BibitemOpen
  \bibfield  {author} {\bibinfo {author} {\bibfnamefont {L.}~\bibnamefont
  {Sarma}}, \bibinfo {author} {\bibfnamefont {P.}~\bibnamefont {Das}}, \ and\
  \bibinfo {author} {\bibfnamefont {M.~K.}\ \bibnamefont {Das}},\ }\href
  {\doibase 10.1016/j.nuclphysb.2020.115300} {\bibfield  {journal} {\bibinfo
  {journal} {Nucl. Phys. B}\ }\textbf {\bibinfo {volume} {963}},\ \bibinfo
  {pages} {115300} (\bibinfo {year} {2021})},\ \Eprint
  {http://arxiv.org/abs/2004.13762} {arXiv:2004.13762 [hep-ph]} \BibitemShut
  {NoStop}%
\bibitem [{\citenamefont {Aprile}\ \emph {et~al.}(2018)\citenamefont {Aprile}
  \emph {et~al.}}]{Aprile:2018dbl}%
  \BibitemOpen
  \bibfield  {author} {\bibinfo {author} {\bibfnamefont {E.}~\bibnamefont
  {Aprile}} \emph {et~al.} (\bibinfo {collaboration} {XENON}),\ }\href
  {\doibase 10.1103/PhysRevLett.121.111302} {\bibfield  {journal} {\bibinfo
  {journal} {Phys. Rev. Lett.}\ }\textbf {\bibinfo {volume} {121}},\ \bibinfo
  {pages} {111302} (\bibinfo {year} {2018})},\ \Eprint
  {http://arxiv.org/abs/1805.12562} {arXiv:1805.12562 [astro-ph.CO]}
  \BibitemShut {NoStop}%
\bibitem [{\citenamefont {Deshpande}\ and\ \citenamefont
  {Ma}(1978)}]{Deshpande:1977rw}%
  \BibitemOpen
  \bibfield  {author} {\bibinfo {author} {\bibfnamefont {N.~G.}\ \bibnamefont
  {Deshpande}}\ and\ \bibinfo {author} {\bibfnamefont {E.}~\bibnamefont {Ma}},\
  }\href {\doibase 10.1103/PhysRevD.18.2574} {\bibfield  {journal} {\bibinfo
  {journal} {Phys. Rev.}\ }\textbf {\bibinfo {volume} {D18}},\ \bibinfo {pages}
  {2574} (\bibinfo {year} {1978})}\BibitemShut {NoStop}%
\bibitem [{\citenamefont {Khan}(2018)}]{Khan:2016sxm}%
  \BibitemOpen
  \bibfield  {author} {\bibinfo {author} {\bibfnamefont {N.}~\bibnamefont
  {Khan}},\ }\href {\doibase 10.1140/epjc/s10052-018-5766-4} {\bibfield
  {journal} {\bibinfo  {journal} {Eur. Phys. J.}\ }\textbf {\bibinfo {volume}
  {C78}},\ \bibinfo {pages} {341} (\bibinfo {year} {2018})},\ \Eprint
  {http://arxiv.org/abs/1610.03178} {arXiv:1610.03178 [hep-ph]} \BibitemShut
  {NoStop}%
\bibitem [{\citenamefont {Chaudhuri}\ \emph {et~al.}(2015)\citenamefont
  {Chaudhuri}, \citenamefont {Khan}, \citenamefont {Mukhopadhyaya},\ and\
  \citenamefont {Rakshit}}]{Chaudhuri:2015pna}%
  \BibitemOpen
  \bibfield  {author} {\bibinfo {author} {\bibfnamefont {A.}~\bibnamefont
  {Chaudhuri}}, \bibinfo {author} {\bibfnamefont {N.}~\bibnamefont {Khan}},
  \bibinfo {author} {\bibfnamefont {B.}~\bibnamefont {Mukhopadhyaya}}, \ and\
  \bibinfo {author} {\bibfnamefont {S.}~\bibnamefont {Rakshit}},\ }\href
  {\doibase 10.1103/PhysRevD.91.055024} {\bibfield  {journal} {\bibinfo
  {journal} {Phys. Rev. D}\ }\textbf {\bibinfo {volume} {91}},\ \bibinfo
  {pages} {055024} (\bibinfo {year} {2015})},\ \Eprint
  {http://arxiv.org/abs/1501.05885} {arXiv:1501.05885 [hep-ph]} \BibitemShut
  {NoStop}%
\bibitem [{\citenamefont {Dienes}\ and\ \citenamefont
  {Thomas}(2012)}]{Dienes:2011ja}%
  \BibitemOpen
  \bibfield  {author} {\bibinfo {author} {\bibfnamefont {K.~R.}\ \bibnamefont
  {Dienes}}\ and\ \bibinfo {author} {\bibfnamefont {B.}~\bibnamefont
  {Thomas}},\ }\href {\doibase 10.1103/PhysRevD.85.083523} {\bibfield
  {journal} {\bibinfo  {journal} {Phys. Rev. D}\ }\textbf {\bibinfo {volume}
  {85}},\ \bibinfo {pages} {083523} (\bibinfo {year} {2012})},\ \Eprint
  {http://arxiv.org/abs/1106.4546} {arXiv:1106.4546 [hep-ph]} \BibitemShut
  {NoStop}%
\bibitem [{\citenamefont {Biswas}\ and\ \citenamefont
  {Gupta}(2016)}]{Biswas:2016bfo}%
  \BibitemOpen
  \bibfield  {author} {\bibinfo {author} {\bibfnamefont {A.}~\bibnamefont
  {Biswas}}\ and\ \bibinfo {author} {\bibfnamefont {A.}~\bibnamefont {Gupta}},\
  }\href {\doibase 10.1088/1475-7516/2016/09/044} {\bibfield  {journal}
  {\bibinfo  {journal} {JCAP}\ }\textbf {\bibinfo {volume} {09}},\ \bibinfo
  {pages} {044} (\bibinfo {year} {2016})},\ \bibinfo {note} {[Addendum: JCAP
  05, A01 (2017)]},\ \Eprint {http://arxiv.org/abs/1607.01469}
  {arXiv:1607.01469 [hep-ph]} \BibitemShut {NoStop}%
\bibitem [{\citenamefont {Elahi}\ \emph {et~al.}(2015)\citenamefont {Elahi},
  \citenamefont {Kolda},\ and\ \citenamefont {Unwin}}]{Elahi:2014fsa}%
  \BibitemOpen
  \bibfield  {author} {\bibinfo {author} {\bibfnamefont {F.}~\bibnamefont
  {Elahi}}, \bibinfo {author} {\bibfnamefont {C.}~\bibnamefont {Kolda}}, \ and\
  \bibinfo {author} {\bibfnamefont {J.}~\bibnamefont {Unwin}},\ }\href
  {\doibase 10.1007/JHEP03(2015)048} {\bibfield  {journal} {\bibinfo  {journal}
  {JHEP}\ }\textbf {\bibinfo {volume} {03}},\ \bibinfo {pages} {048} (\bibinfo
  {year} {2015})},\ \Eprint {http://arxiv.org/abs/1410.6157} {arXiv:1410.6157
  [hep-ph]} \BibitemShut {NoStop}%
\bibitem [{\citenamefont {Abi}\ \emph {et~al.}(2021)\citenamefont {Abi} \emph
  {et~al.}}]{Abi:2021gix}%
  \BibitemOpen
  \bibfield  {author} {\bibinfo {author} {\bibfnamefont {B.}~\bibnamefont
  {Abi}} \emph {et~al.} (\bibinfo {collaboration} {Muon g-2}),\ }\href
  {\doibase 10.1103/PhysRevLett.126.141801} {\bibfield  {journal} {\bibinfo
  {journal} {Phys. Rev. Lett.}\ }\textbf {\bibinfo {volume} {126}},\ \bibinfo
  {pages} {141801} (\bibinfo {year} {2021})},\ \Eprint
  {http://arxiv.org/abs/2104.03281} {arXiv:2104.03281 [hep-ex]} \BibitemShut
  {NoStop}%
\bibitem [{\citenamefont {Bennett}\ \emph
  {et~al.}(2006{\natexlab{a}})\citenamefont {Bennett} \emph
  {et~al.}}]{Bennett:2006fi}%
  \BibitemOpen
  \bibfield  {author} {\bibinfo {author} {\bibfnamefont {G.~W.}\ \bibnamefont
  {Bennett}} \emph {et~al.} (\bibinfo {collaboration} {Muon g-2}),\ }\href
  {\doibase 10.1103/PhysRevD.73.072003} {\bibfield  {journal} {\bibinfo
  {journal} {Phys. Rev. D}\ }\textbf {\bibinfo {volume} {73}},\ \bibinfo
  {pages} {072003} (\bibinfo {year} {2006}{\natexlab{a}})},\ \Eprint
  {http://arxiv.org/abs/hep-ex/0602035} {arXiv:hep-ex/0602035} \BibitemShut
  {NoStop}%
\bibitem [{\citenamefont {Parker}\ \emph {et~al.}(2018)\citenamefont {Parker},
  \citenamefont {Yu}, \citenamefont {Zhong}, \citenamefont {Estey},\ and\
  \citenamefont {M\"uller}}]{Parker:2018vye}%
  \BibitemOpen
  \bibfield  {author} {\bibinfo {author} {\bibfnamefont {R.~H.}\ \bibnamefont
  {Parker}}, \bibinfo {author} {\bibfnamefont {C.}~\bibnamefont {Yu}}, \bibinfo
  {author} {\bibfnamefont {W.}~\bibnamefont {Zhong}}, \bibinfo {author}
  {\bibfnamefont {B.}~\bibnamefont {Estey}}, \ and\ \bibinfo {author}
  {\bibfnamefont {H.}~\bibnamefont {M\"uller}},\ }\href {\doibase
  10.1126/science.aap7706} {\bibfield  {journal} {\bibinfo  {journal}
  {Science}\ }\textbf {\bibinfo {volume} {360}},\ \bibinfo {pages} {191}
  (\bibinfo {year} {2018})},\ \Eprint {http://arxiv.org/abs/1812.04130}
  {arXiv:1812.04130 [physics.atom-ph]} \BibitemShut {NoStop}%
\bibitem [{\citenamefont {Bennett}\ \emph
  {et~al.}(2006{\natexlab{b}})\citenamefont {Bennett} \emph
  {et~al.}}]{Muong-2:2006rrc}%
  \BibitemOpen
  \bibfield  {author} {\bibinfo {author} {\bibfnamefont {G.~W.}\ \bibnamefont
  {Bennett}} \emph {et~al.} (\bibinfo {collaboration} {Muon g-2}),\ }\href
  {\doibase 10.1103/PhysRevD.73.072003} {\bibfield  {journal} {\bibinfo
  {journal} {Phys. Rev. D}\ }\textbf {\bibinfo {volume} {73}},\ \bibinfo
  {pages} {072003} (\bibinfo {year} {2006}{\natexlab{b}})},\ \Eprint
  {http://arxiv.org/abs/hep-ex/0602035} {arXiv:hep-ex/0602035} \BibitemShut
  {NoStop}%
\bibitem [{\citenamefont {Borah}\ \emph {et~al.}(2021)\citenamefont {Borah},
  \citenamefont {Dutta}, \citenamefont {Mahapatra},\ and\ \citenamefont
  {Sahu}}]{Borah:2021jzu}%
  \BibitemOpen
  \bibfield  {author} {\bibinfo {author} {\bibfnamefont {D.}~\bibnamefont
  {Borah}}, \bibinfo {author} {\bibfnamefont {M.}~\bibnamefont {Dutta}},
  \bibinfo {author} {\bibfnamefont {S.}~\bibnamefont {Mahapatra}}, \ and\
  \bibinfo {author} {\bibfnamefont {N.}~\bibnamefont {Sahu}},\ }\href@noop {}
  {\  (\bibinfo {year} {2021})},\ \Eprint {http://arxiv.org/abs/2104.05656}
  {arXiv:2104.05656 [hep-ph]} \BibitemShut {NoStop}%
\bibitem [{\citenamefont {Dasgupta}\ \emph {et~al.}(2021)\citenamefont
  {Dasgupta}, \citenamefont {Kang},\ and\ \citenamefont
  {Park}}]{Dasgupta:2021dnl}%
  \BibitemOpen
  \bibfield  {author} {\bibinfo {author} {\bibfnamefont {A.}~\bibnamefont
  {Dasgupta}}, \bibinfo {author} {\bibfnamefont {S.~K.}\ \bibnamefont {Kang}},
  \ and\ \bibinfo {author} {\bibfnamefont {M.}~\bibnamefont {Park}},\
  }\href@noop {} {\  (\bibinfo {year} {2021})},\ \Eprint
  {http://arxiv.org/abs/2104.09205} {arXiv:2104.09205 [hep-ph]} \BibitemShut
  {NoStop}%
\bibitem [{\citenamefont {Cen}\ \emph {et~al.}(2021)\citenamefont {Cen},
  \citenamefont {Cheng}, \citenamefont {He},\ and\ \citenamefont
  {Sun}}]{Cen:2021iwv}%
  \BibitemOpen
  \bibfield  {author} {\bibinfo {author} {\bibfnamefont {J.-Y.}\ \bibnamefont
  {Cen}}, \bibinfo {author} {\bibfnamefont {Y.}~\bibnamefont {Cheng}}, \bibinfo
  {author} {\bibfnamefont {X.-G.}\ \bibnamefont {He}}, \ and\ \bibinfo {author}
  {\bibfnamefont {J.}~\bibnamefont {Sun}},\ }\href@noop {} {\  (\bibinfo {year}
  {2021})},\ \Eprint {http://arxiv.org/abs/2104.05006} {arXiv:2104.05006
  [hep-ph]} \BibitemShut {NoStop}%
\bibitem [{\citenamefont {Yang}\ \emph {et~al.}(2021)\citenamefont {Yang},
  \citenamefont {Zhang}, \citenamefont {Liu}, \citenamefont {Dong},\ and\
  \citenamefont {Feng}}]{Yang:2021duj}%
  \BibitemOpen
  \bibfield  {author} {\bibinfo {author} {\bibfnamefont {J.-L.}\ \bibnamefont
  {Yang}}, \bibinfo {author} {\bibfnamefont {H.-B.}\ \bibnamefont {Zhang}},
  \bibinfo {author} {\bibfnamefont {C.-X.}\ \bibnamefont {Liu}}, \bibinfo
  {author} {\bibfnamefont {X.-X.}\ \bibnamefont {Dong}}, \ and\ \bibinfo
  {author} {\bibfnamefont {T.-F.}\ \bibnamefont {Feng}},\ }\href@noop {} {\
  (\bibinfo {year} {2021})},\ \Eprint {http://arxiv.org/abs/2104.03542}
  {arXiv:2104.03542 [hep-ph]} \BibitemShut {NoStop}%
\bibitem [{\citenamefont {Das}\ \emph {et~al.}(2021{\natexlab{b}})\citenamefont
  {Das}, \citenamefont {Das},\ and\ \citenamefont {Khan}}]{Das:2021zea}%
  \BibitemOpen
  \bibfield  {author} {\bibinfo {author} {\bibfnamefont {P.}~\bibnamefont
  {Das}}, \bibinfo {author} {\bibfnamefont {M.~K.}\ \bibnamefont {Das}}, \ and\
  \bibinfo {author} {\bibfnamefont {N.}~\bibnamefont {Khan}},\ }\href@noop {}
  {\  (\bibinfo {year} {2021}{\natexlab{b}})},\ \Eprint
  {http://arxiv.org/abs/2104.03271} {arXiv:2104.03271 [hep-ph]} \BibitemShut
  {NoStop}%
\bibitem [{\citenamefont {Escribano}\ \emph {et~al.}(2021)\citenamefont
  {Escribano}, \citenamefont {Terol-Calvo},\ and\ \citenamefont
  {Vicente}}]{Escribano:2021css}%
  \BibitemOpen
  \bibfield  {author} {\bibinfo {author} {\bibfnamefont {P.}~\bibnamefont
  {Escribano}}, \bibinfo {author} {\bibfnamefont {J.}~\bibnamefont
  {Terol-Calvo}}, \ and\ \bibinfo {author} {\bibfnamefont {A.}~\bibnamefont
  {Vicente}},\ }\href@noop {} {\  (\bibinfo {year} {2021})},\ \Eprint
  {http://arxiv.org/abs/2104.03705} {arXiv:2104.03705 [hep-ph]} \BibitemShut
  {NoStop}%
\bibitem [{\citenamefont {Athron}\ \emph {et~al.}(2021)\citenamefont {Athron},
  \citenamefont {Bal\'azs}, \citenamefont {Jacob}, \citenamefont {Kotlarski},
  \citenamefont {St\"ockinger},\ and\ \citenamefont
  {St\"ockinger-Kim}}]{Athron:2021iuf}%
  \BibitemOpen
  \bibfield  {author} {\bibinfo {author} {\bibfnamefont {P.}~\bibnamefont
  {Athron}}, \bibinfo {author} {\bibfnamefont {C.}~\bibnamefont {Bal\'azs}},
  \bibinfo {author} {\bibfnamefont {D.~H.}\ \bibnamefont {Jacob}}, \bibinfo
  {author} {\bibfnamefont {W.}~\bibnamefont {Kotlarski}}, \bibinfo {author}
  {\bibfnamefont {D.}~\bibnamefont {St\"ockinger}}, \ and\ \bibinfo {author}
  {\bibfnamefont {H.}~\bibnamefont {St\"ockinger-Kim}},\ }\href@noop {} {\
  (\bibinfo {year} {2021})},\ \Eprint {http://arxiv.org/abs/2104.03691}
  {arXiv:2104.03691 [hep-ph]} \BibitemShut {NoStop}%
\bibitem [{\citenamefont {Kowalska}\ and\ \citenamefont
  {Sessolo}(2017)}]{Kowalska:2017iqv}%
  \BibitemOpen
  \bibfield  {author} {\bibinfo {author} {\bibfnamefont {K.}~\bibnamefont
  {Kowalska}}\ and\ \bibinfo {author} {\bibfnamefont {E.~M.}\ \bibnamefont
  {Sessolo}},\ }\href {\doibase 10.1007/JHEP09(2017)112} {\bibfield  {journal}
  {\bibinfo  {journal} {JHEP}\ }\textbf {\bibinfo {volume} {09}},\ \bibinfo
  {pages} {112} (\bibinfo {year} {2017})},\ \Eprint
  {http://arxiv.org/abs/1707.00753} {arXiv:1707.00753 [hep-ph]} \BibitemShut
  {NoStop}%
\bibitem [{\citenamefont {Calibbi}\ \emph {et~al.}(2018)\citenamefont
  {Calibbi}, \citenamefont {Ziegler},\ and\ \citenamefont
  {Zupan}}]{Calibbi:2018rzv}%
  \BibitemOpen
  \bibfield  {author} {\bibinfo {author} {\bibfnamefont {L.}~\bibnamefont
  {Calibbi}}, \bibinfo {author} {\bibfnamefont {R.}~\bibnamefont {Ziegler}}, \
  and\ \bibinfo {author} {\bibfnamefont {J.}~\bibnamefont {Zupan}},\ }\href
  {\doibase 10.1007/JHEP07(2018)046} {\bibfield  {journal} {\bibinfo  {journal}
  {JHEP}\ }\textbf {\bibinfo {volume} {07}},\ \bibinfo {pages} {046} (\bibinfo
  {year} {2018})},\ \Eprint {http://arxiv.org/abs/1804.00009} {arXiv:1804.00009
  [hep-ph]} \BibitemShut {NoStop}%
\bibitem [{\citenamefont {Kowalska}\ and\ \citenamefont
  {Sessolo}(2020)}]{Kowalska:2020zve}%
  \BibitemOpen
  \bibfield  {author} {\bibinfo {author} {\bibfnamefont {K.}~\bibnamefont
  {Kowalska}}\ and\ \bibinfo {author} {\bibfnamefont {E.~M.}\ \bibnamefont
  {Sessolo}},\ }\href@noop {} {\  (\bibinfo {year} {2020})},\ \Eprint
  {http://arxiv.org/abs/2012.15200} {arXiv:2012.15200 [hep-ph]} \BibitemShut
  {NoStop}%
\bibitem [{\citenamefont {Kawamura}\ \emph {et~al.}(2020)\citenamefont
  {Kawamura}, \citenamefont {Okawa},\ and\ \citenamefont
  {Omura}}]{Kawamura:2020qxo}%
  \BibitemOpen
  \bibfield  {author} {\bibinfo {author} {\bibfnamefont {J.}~\bibnamefont
  {Kawamura}}, \bibinfo {author} {\bibfnamefont {S.}~\bibnamefont {Okawa}}, \
  and\ \bibinfo {author} {\bibfnamefont {Y.}~\bibnamefont {Omura}},\ }\href
  {\doibase 10.1007/JHEP08(2020)042} {\bibfield  {journal} {\bibinfo  {journal}
  {JHEP}\ }\textbf {\bibinfo {volume} {08}},\ \bibinfo {pages} {042} (\bibinfo
  {year} {2020})},\ \Eprint {http://arxiv.org/abs/2002.12534} {arXiv:2002.12534
  [hep-ph]} \BibitemShut {NoStop}%
\bibitem [{\citenamefont {Baker}\ \emph {et~al.}(2021)\citenamefont {Baker},
  \citenamefont {Cox},\ and\ \citenamefont {Volkas}}]{Baker:2021yli}%
  \BibitemOpen
  \bibfield  {author} {\bibinfo {author} {\bibfnamefont {M.~J.}\ \bibnamefont
  {Baker}}, \bibinfo {author} {\bibfnamefont {P.}~\bibnamefont {Cox}}, \ and\
  \bibinfo {author} {\bibfnamefont {R.~R.}\ \bibnamefont {Volkas}},\ }\href
  {\doibase 10.1007/JHEP05(2021)174} {\bibfield  {journal} {\bibinfo  {journal}
  {JHEP}\ }\textbf {\bibinfo {volume} {05}},\ \bibinfo {pages} {174} (\bibinfo
  {year} {2021})},\ \Eprint {http://arxiv.org/abs/2103.13401} {arXiv:2103.13401
  [hep-ph]} \BibitemShut {NoStop}%
\bibitem [{\citenamefont {G\'erardin}\ \emph {et~al.}(2019)\citenamefont
  {G\'erardin}, \citenamefont {C\`e}, \citenamefont {von Hippel}, \citenamefont
  {H\"orz}, \citenamefont {Meyer}, \citenamefont {Mohler}, \citenamefont
  {Ottnad}, \citenamefont {Wilhelm},\ and\ \citenamefont
  {Wittig}}]{Gerardin:2019rua}%
  \BibitemOpen
  \bibfield  {author} {\bibinfo {author} {\bibfnamefont {A.}~\bibnamefont
  {G\'erardin}}, \bibinfo {author} {\bibfnamefont {M.}~\bibnamefont {C\`e}},
  \bibinfo {author} {\bibfnamefont {G.}~\bibnamefont {von Hippel}}, \bibinfo
  {author} {\bibfnamefont {B.}~\bibnamefont {H\"orz}}, \bibinfo {author}
  {\bibfnamefont {H.~B.}\ \bibnamefont {Meyer}}, \bibinfo {author}
  {\bibfnamefont {D.}~\bibnamefont {Mohler}}, \bibinfo {author} {\bibfnamefont
  {K.}~\bibnamefont {Ottnad}}, \bibinfo {author} {\bibfnamefont
  {J.}~\bibnamefont {Wilhelm}}, \ and\ \bibinfo {author} {\bibfnamefont
  {H.}~\bibnamefont {Wittig}},\ }\href {\doibase 10.1103/PhysRevD.100.014510}
  {\bibfield  {journal} {\bibinfo  {journal} {Phys. Rev. D}\ }\textbf {\bibinfo
  {volume} {100}},\ \bibinfo {pages} {014510} (\bibinfo {year} {2019})},\
  \Eprint {http://arxiv.org/abs/1904.03120} {arXiv:1904.03120 [hep-lat]}
  \BibitemShut {NoStop}%
\bibitem [{\citenamefont {G\'erardin}(2021)}]{Gerardin:2020gpp}%
  \BibitemOpen
  \bibfield  {author} {\bibinfo {author} {\bibfnamefont {A.}~\bibnamefont
  {G\'erardin}},\ }\href {\doibase 10.1140/epja/s10050-021-00426-7} {\bibfield
  {journal} {\bibinfo  {journal} {Eur. Phys. J. A}\ }\textbf {\bibinfo {volume}
  {57}},\ \bibinfo {pages} {116} (\bibinfo {year} {2021})},\ \Eprint
  {http://arxiv.org/abs/2012.03931} {arXiv:2012.03931 [hep-lat]} \BibitemShut
  {NoStop}%
\bibitem [{\citenamefont {Chao}\ \emph {et~al.}(2021)\citenamefont {Chao},
  \citenamefont {Hudspith}, \citenamefont {G\'erardin}, \citenamefont {Green},
  \citenamefont {Meyer},\ and\ \citenamefont {Ottnad}}]{Chao:2021tvp}%
  \BibitemOpen
  \bibfield  {author} {\bibinfo {author} {\bibfnamefont {E.-H.}\ \bibnamefont
  {Chao}}, \bibinfo {author} {\bibfnamefont {R.~J.}\ \bibnamefont {Hudspith}},
  \bibinfo {author} {\bibfnamefont {A.}~\bibnamefont {G\'erardin}}, \bibinfo
  {author} {\bibfnamefont {J.~R.}\ \bibnamefont {Green}}, \bibinfo {author}
  {\bibfnamefont {H.~B.}\ \bibnamefont {Meyer}}, \ and\ \bibinfo {author}
  {\bibfnamefont {K.}~\bibnamefont {Ottnad}},\ }\href {\doibase
  10.1140/epjc/s10052-021-09455-4} {\bibfield  {journal} {\bibinfo  {journal}
  {Eur. Phys. J. C}\ }\textbf {\bibinfo {volume} {81}},\ \bibinfo {pages} {651}
  (\bibinfo {year} {2021})},\ \Eprint {http://arxiv.org/abs/2104.02632}
  {arXiv:2104.02632 [hep-lat]} \BibitemShut {NoStop}%
\bibitem [{\citenamefont {Davies}\ \emph {et~al.}(2020)\citenamefont {Davies}
  \emph {et~al.}}]{FermilabLattice:2019ugu}%
  \BibitemOpen
  \bibfield  {author} {\bibinfo {author} {\bibfnamefont {C.~T.~H.}\
  \bibnamefont {Davies}} \emph {et~al.} (\bibinfo {collaboration} {Fermilab
  Lattice, LATTICE-HPQCD, MILC}),\ }\href {\doibase
  10.1103/PhysRevD.101.034512} {\bibfield  {journal} {\bibinfo  {journal}
  {Phys. Rev. D}\ }\textbf {\bibinfo {volume} {101}},\ \bibinfo {pages}
  {034512} (\bibinfo {year} {2020})},\ \Eprint
  {http://arxiv.org/abs/1902.04223} {arXiv:1902.04223 [hep-lat]} \BibitemShut
  {NoStop}%
\bibitem [{\citenamefont {Blank}\ and\ \citenamefont
  {Hollik}(1998)}]{Blank:1997qa}%
  \BibitemOpen
  \bibfield  {author} {\bibinfo {author} {\bibfnamefont {T.}~\bibnamefont
  {Blank}}\ and\ \bibinfo {author} {\bibfnamefont {W.}~\bibnamefont {Hollik}},\
  }\href {\doibase 10.1016/S0550-3213(97)00785-2} {\bibfield  {journal}
  {\bibinfo  {journal} {Nucl. Phys. B}\ }\textbf {\bibinfo {volume} {514}},\
  \bibinfo {pages} {113} (\bibinfo {year} {1998})},\ \Eprint
  {http://arxiv.org/abs/hep-ph/9703392} {arXiv:hep-ph/9703392} \BibitemShut
  {NoStop}%
\bibitem [{\citenamefont {Pal}(2014)}]{Pal:1690642}%
  \BibitemOpen
  \bibfield  {author} {\bibinfo {author} {\bibfnamefont {P.~B.}\ \bibnamefont
  {Pal}},\ }\href {\doibase 1482216981} {\emph {\bibinfo {title} {{An
  introductory course of particle physics}}}}\ (\bibinfo {address} {Boca Raton,
  FL},\ \bibinfo {year} {2014})\BibitemShut {NoStop}%
\bibitem [{\citenamefont {Chen}\ \emph {et~al.}(2008)\citenamefont {Chen},
  \citenamefont {Dawson},\ and\ \citenamefont {Jackson}}]{Chen:2008jg}%
  \BibitemOpen
  \bibfield  {author} {\bibinfo {author} {\bibfnamefont {M.-C.}\ \bibnamefont
  {Chen}}, \bibinfo {author} {\bibfnamefont {S.}~\bibnamefont {Dawson}}, \ and\
  \bibinfo {author} {\bibfnamefont {C.~B.}\ \bibnamefont {Jackson}},\ }\href
  {\doibase 10.1103/PhysRevD.78.093001} {\bibfield  {journal} {\bibinfo
  {journal} {Phys. Rev. D}\ }\textbf {\bibinfo {volume} {78}},\ \bibinfo
  {pages} {093001} (\bibinfo {year} {2008})},\ \Eprint
  {http://arxiv.org/abs/0809.4185} {arXiv:0809.4185 [hep-ph]} \BibitemShut
  {NoStop}%
\bibitem [{\citenamefont {Fiaschi}\ \emph {et~al.}(2019)\citenamefont
  {Fiaschi}, \citenamefont {Klasen},\ and\ \citenamefont
  {May}}]{Fiaschi:2018rky}%
  \BibitemOpen
  \bibfield  {author} {\bibinfo {author} {\bibfnamefont {J.}~\bibnamefont
  {Fiaschi}}, \bibinfo {author} {\bibfnamefont {M.}~\bibnamefont {Klasen}}, \
  and\ \bibinfo {author} {\bibfnamefont {S.}~\bibnamefont {May}},\ }\href
  {\doibase 10.1007/JHEP05(2019)015} {\bibfield  {journal} {\bibinfo  {journal}
  {JHEP}\ }\textbf {\bibinfo {volume} {05}},\ \bibinfo {pages} {015} (\bibinfo
  {year} {2019})},\ \Eprint {http://arxiv.org/abs/1812.11133} {arXiv:1812.11133
  [hep-ph]} \BibitemShut {NoStop}%
\bibitem [{\citenamefont {Cirelli}\ and\ \citenamefont
  {Strumia}(2009)}]{Cirelli:2009uv}%
  \BibitemOpen
  \bibfield  {author} {\bibinfo {author} {\bibfnamefont {M.}~\bibnamefont
  {Cirelli}}\ and\ \bibinfo {author} {\bibfnamefont {A.}~\bibnamefont
  {Strumia}},\ }\href {\doibase 10.1088/1367-2630/11/10/105005} {\bibfield
  {journal} {\bibinfo  {journal} {New J. Phys.}\ }\textbf {\bibinfo {volume}
  {11}},\ \bibinfo {pages} {105005} (\bibinfo {year} {2009})},\ \Eprint
  {http://arxiv.org/abs/0903.3381} {arXiv:0903.3381 [hep-ph]} \BibitemShut
  {NoStop}%
\bibitem [{\citenamefont {Cirelli}\ \emph {et~al.}(2006)\citenamefont
  {Cirelli}, \citenamefont {Fornengo},\ and\ \citenamefont
  {Strumia}}]{Cirelli:2005uq}%
  \BibitemOpen
  \bibfield  {author} {\bibinfo {author} {\bibfnamefont {M.}~\bibnamefont
  {Cirelli}}, \bibinfo {author} {\bibfnamefont {N.}~\bibnamefont {Fornengo}}, \
  and\ \bibinfo {author} {\bibfnamefont {A.}~\bibnamefont {Strumia}},\ }\href
  {\doibase 10.1016/j.nuclphysb.2006.07.012} {\bibfield  {journal} {\bibinfo
  {journal} {Nucl. Phys.}\ }\textbf {\bibinfo {volume} {B753}},\ \bibinfo
  {pages} {178} (\bibinfo {year} {2006})},\ \Eprint
  {http://arxiv.org/abs/hep-ph/0512090} {arXiv:hep-ph/0512090 [hep-ph]}
  \BibitemShut {NoStop}%
\bibitem [{\citenamefont {Kannike}(2012)}]{Kannike:2012pe}%
  \BibitemOpen
  \bibfield  {author} {\bibinfo {author} {\bibfnamefont {K.}~\bibnamefont
  {Kannike}},\ }\href {\doibase 10.1140/epjc/s10052-012-2093-z} {\bibfield
  {journal} {\bibinfo  {journal} {Eur. Phys. J. C}\ }\textbf {\bibinfo {volume}
  {72}},\ \bibinfo {pages} {2093} (\bibinfo {year} {2012})},\ \Eprint
  {http://arxiv.org/abs/1205.3781} {arXiv:1205.3781 [hep-ph]} \BibitemShut
  {NoStop}%
\bibitem [{\citenamefont {Peskin}\ and\ \citenamefont
  {Takeuchi}(1992)}]{Peskin:1991sw}%
  \BibitemOpen
  \bibfield  {author} {\bibinfo {author} {\bibfnamefont {M.~E.}\ \bibnamefont
  {Peskin}}\ and\ \bibinfo {author} {\bibfnamefont {T.}~\bibnamefont
  {Takeuchi}},\ }\href {\doibase 10.1103/PhysRevD.46.381} {\bibfield  {journal}
  {\bibinfo  {journal} {Phys. Rev.}\ }\textbf {\bibinfo {volume} {D46}},\
  \bibinfo {pages} {381} (\bibinfo {year} {1992})}\BibitemShut {NoStop}%
\bibitem [{\citenamefont {Forshaw}\ \emph {et~al.}(2003)\citenamefont
  {Forshaw}, \citenamefont {Sabio~Vera},\ and\ \citenamefont
  {White}}]{Forshaw:2003kh}%
  \BibitemOpen
  \bibfield  {author} {\bibinfo {author} {\bibfnamefont {J.~R.}\ \bibnamefont
  {Forshaw}}, \bibinfo {author} {\bibfnamefont {A.}~\bibnamefont {Sabio~Vera}},
  \ and\ \bibinfo {author} {\bibfnamefont {B.~E.}\ \bibnamefont {White}},\
  }\href {\doibase 10.1088/1126-6708/2003/06/059} {\bibfield  {journal}
  {\bibinfo  {journal} {JHEP}\ }\textbf {\bibinfo {volume} {06}},\ \bibinfo
  {pages} {059} (\bibinfo {year} {2003})},\ \Eprint
  {http://arxiv.org/abs/hep-ph/0302256} {arXiv:hep-ph/0302256} \BibitemShut
  {NoStop}%
\bibitem [{\citenamefont {Forshaw}\ \emph {et~al.}(2001)\citenamefont
  {Forshaw}, \citenamefont {Ross},\ and\ \citenamefont
  {White}}]{Forshaw:2001xq}%
  \BibitemOpen
  \bibfield  {author} {\bibinfo {author} {\bibfnamefont {J.~R.}\ \bibnamefont
  {Forshaw}}, \bibinfo {author} {\bibfnamefont {D.~A.}\ \bibnamefont {Ross}}, \
  and\ \bibinfo {author} {\bibfnamefont {B.~E.}\ \bibnamefont {White}},\ }\href
  {\doibase 10.1088/1126-6708/2001/10/007} {\bibfield  {journal} {\bibinfo
  {journal} {JHEP}\ }\textbf {\bibinfo {volume} {10}},\ \bibinfo {pages} {007}
  (\bibinfo {year} {2001})},\ \Eprint {http://arxiv.org/abs/hep-ph/0107232}
  {arXiv:hep-ph/0107232} \BibitemShut {NoStop}%
\bibitem [{\citenamefont {Baak}\ \emph {et~al.}(2014)\citenamefont {Baak},
  \citenamefont {Cúth}, \citenamefont {Haller}, \citenamefont {Hoecker},
  \citenamefont {Kogler}, \citenamefont {Mönig}, \citenamefont {Schott},\ and\
  \citenamefont {Stelzer}}]{Baak:2014ora}%
  \BibitemOpen
  \bibfield  {author} {\bibinfo {author} {\bibfnamefont {M.}~\bibnamefont
  {Baak}}, \bibinfo {author} {\bibfnamefont {J.}~\bibnamefont {Cúth}},
  \bibinfo {author} {\bibfnamefont {J.}~\bibnamefont {Haller}}, \bibinfo
  {author} {\bibfnamefont {A.}~\bibnamefont {Hoecker}}, \bibinfo {author}
  {\bibfnamefont {R.}~\bibnamefont {Kogler}}, \bibinfo {author} {\bibfnamefont
  {K.}~\bibnamefont {Mönig}}, \bibinfo {author} {\bibfnamefont
  {M.}~\bibnamefont {Schott}}, \ and\ \bibinfo {author} {\bibfnamefont
  {J.}~\bibnamefont {Stelzer}} (\bibinfo {collaboration} {Gfitter Group}),\
  }\href {\doibase 10.1140/epjc/s10052-014-3046-5} {\bibfield  {journal}
  {\bibinfo  {journal} {Eur. Phys. J.}\ }\textbf {\bibinfo {volume} {C74}},\
  \bibinfo {pages} {3046} (\bibinfo {year} {2014})},\ \Eprint
  {http://arxiv.org/abs/1407.3792} {arXiv:1407.3792 [hep-ph]} \BibitemShut
  {NoStop}%
\bibitem [{\citenamefont {Hieu}\ \emph {et~al.}(2020)\citenamefont {Hieu},
  \citenamefont {Sang},\ and\ \citenamefont {Trang}}]{Hieu:2020hti}%
  \BibitemOpen
  \bibfield  {author} {\bibinfo {author} {\bibfnamefont {T.~M.}\ \bibnamefont
  {Hieu}}, \bibinfo {author} {\bibfnamefont {Q.~S.}\ \bibnamefont {Sang}}, \
  and\ \bibinfo {author} {\bibfnamefont {T.~Q.}\ \bibnamefont {Trang}},\ }\href
  {\doibase 10.15625/0868-3166/30/3/15071} {\bibfield  {journal} {\bibinfo
  {journal} {Commun. in Phys.}\ }\textbf {\bibinfo {volume} {30}},\ \bibinfo
  {pages} {231} (\bibinfo {year} {2020})}\BibitemShut {NoStop}%
\bibitem [{\citenamefont {Belanger}\ \emph {et~al.}(2013)\citenamefont
  {Belanger}, \citenamefont {Dumont}, \citenamefont {Ellwanger}, \citenamefont
  {Gunion},\ and\ \citenamefont {Kraml}}]{Belanger:2013xza}%
  \BibitemOpen
  \bibfield  {author} {\bibinfo {author} {\bibfnamefont {G.}~\bibnamefont
  {Belanger}}, \bibinfo {author} {\bibfnamefont {B.}~\bibnamefont {Dumont}},
  \bibinfo {author} {\bibfnamefont {U.}~\bibnamefont {Ellwanger}}, \bibinfo
  {author} {\bibfnamefont {J.~F.}\ \bibnamefont {Gunion}}, \ and\ \bibinfo
  {author} {\bibfnamefont {S.}~\bibnamefont {Kraml}},\ }\href {\doibase
  10.1103/PhysRevD.88.075008} {\bibfield  {journal} {\bibinfo  {journal} {Phys.
  Rev. D}\ }\textbf {\bibinfo {volume} {88}},\ \bibinfo {pages} {075008}
  (\bibinfo {year} {2013})},\ \Eprint {http://arxiv.org/abs/1306.2941}
  {arXiv:1306.2941 [hep-ph]} \BibitemShut {NoStop}%
\bibitem [{\citenamefont {Djouadi}(2008)}]{Djouadi:2005gj}%
  \BibitemOpen
  \bibfield  {author} {\bibinfo {author} {\bibfnamefont {A.}~\bibnamefont
  {Djouadi}},\ }\href {\doibase 10.1016/j.physrep.2007.10.005} {\bibfield
  {journal} {\bibinfo  {journal} {Phys. Rept.}\ }\textbf {\bibinfo {volume}
  {459}},\ \bibinfo {pages} {1} (\bibinfo {year} {2008})},\ \Eprint
  {http://arxiv.org/abs/hep-ph/0503173} {arXiv:hep-ph/0503173 [hep-ph]}
  \BibitemShut {NoStop}%
\bibitem [{\citenamefont {Aad}\ \emph {et~al.}(2014)\citenamefont {Aad} \emph
  {et~al.}}]{Aad:2014eha}%
  \BibitemOpen
  \bibfield  {author} {\bibinfo {author} {\bibfnamefont {G.}~\bibnamefont
  {Aad}} \emph {et~al.} (\bibinfo {collaboration} {ATLAS}),\ }\href {\doibase
  10.1103/PhysRevD.90.112015} {\bibfield  {journal} {\bibinfo  {journal} {Phys.
  Rev. D}\ }\textbf {\bibinfo {volume} {90}},\ \bibinfo {pages} {112015}
  (\bibinfo {year} {2014})},\ \Eprint {http://arxiv.org/abs/1408.7084}
  {arXiv:1408.7084 [hep-ex]} \BibitemShut {NoStop}%
\bibitem [{\citenamefont {Aaboud}\ \emph {et~al.}(2019)\citenamefont {Aaboud}
  \emph {et~al.}}]{ATLAS:2019cid}%
  \BibitemOpen
  \bibfield  {author} {\bibinfo {author} {\bibfnamefont {M.}~\bibnamefont
  {Aaboud}} \emph {et~al.} (\bibinfo {collaboration} {ATLAS}),\ }\href
  {\doibase 10.1103/PhysRevLett.122.231801} {\bibfield  {journal} {\bibinfo
  {journal} {Phys. Rev. Lett.}\ }\textbf {\bibinfo {volume} {122}},\ \bibinfo
  {pages} {231801} (\bibinfo {year} {2019})},\ \Eprint
  {http://arxiv.org/abs/1904.05105} {arXiv:1904.05105 [hep-ex]} \BibitemShut
  {NoStop}%
\bibitem [{\citenamefont {Khachatryan}\ \emph {et~al.}(2014)\citenamefont
  {Khachatryan} \emph {et~al.}}]{Khachatryan:2014ira}%
  \BibitemOpen
  \bibfield  {author} {\bibinfo {author} {\bibfnamefont {V.}~\bibnamefont
  {Khachatryan}} \emph {et~al.} (\bibinfo {collaboration} {CMS}),\ }\href
  {\doibase 10.1140/epjc/s10052-014-3076-z} {\bibfield  {journal} {\bibinfo
  {journal} {Eur. Phys. J. C}\ }\textbf {\bibinfo {volume} {74}},\ \bibinfo
  {pages} {3076} (\bibinfo {year} {2014})},\ \Eprint
  {http://arxiv.org/abs/1407.0558} {arXiv:1407.0558 [hep-ex]} \BibitemShut
  {NoStop}%
\bibitem [{\citenamefont {Sirunyan}\ \emph
  {et~al.}(2019{\natexlab{b}})\citenamefont {Sirunyan} \emph
  {et~al.}}]{CMS:2018uag}%
  \BibitemOpen
  \bibfield  {author} {\bibinfo {author} {\bibfnamefont {A.~M.}\ \bibnamefont
  {Sirunyan}} \emph {et~al.} (\bibinfo {collaboration} {CMS}),\ }\href
  {\doibase 10.1140/epjc/s10052-019-6909-y} {\bibfield  {journal} {\bibinfo
  {journal} {Eur. Phys. J. C}\ }\textbf {\bibinfo {volume} {79}},\ \bibinfo
  {pages} {421} (\bibinfo {year} {2019}{\natexlab{b}})},\ \Eprint
  {http://arxiv.org/abs/1809.10733} {arXiv:1809.10733 [hep-ex]} \BibitemShut
  {NoStop}%
\bibitem [{\citenamefont {Aad}\ \emph {et~al.}(2016)\citenamefont {Aad} \emph
  {et~al.}}]{Khachatryan:2016vau}%
  \BibitemOpen
  \bibfield  {author} {\bibinfo {author} {\bibfnamefont {G.}~\bibnamefont
  {Aad}} \emph {et~al.} (\bibinfo {collaboration} {ATLAS, CMS}),\ }\href
  {\doibase 10.1007/JHEP08(2016)045} {\bibfield  {journal} {\bibinfo  {journal}
  {JHEP}\ }\textbf {\bibinfo {volume} {08}},\ \bibinfo {pages} {045} (\bibinfo
  {year} {2016})},\ \Eprint {http://arxiv.org/abs/1606.02266} {arXiv:1606.02266
  [hep-ex]} \BibitemShut {NoStop}%
\bibitem [{\citenamefont {Akeroyd}\ \emph {et~al.}(2009)\citenamefont
  {Akeroyd}, \citenamefont {Aoki},\ and\ \citenamefont
  {Sugiyama}}]{Akeroyd:2009nu}%
  \BibitemOpen
  \bibfield  {author} {\bibinfo {author} {\bibfnamefont {A.~G.}\ \bibnamefont
  {Akeroyd}}, \bibinfo {author} {\bibfnamefont {M.}~\bibnamefont {Aoki}}, \
  and\ \bibinfo {author} {\bibfnamefont {H.}~\bibnamefont {Sugiyama}},\ }\href
  {\doibase 10.1103/PhysRevD.79.113010} {\bibfield  {journal} {\bibinfo
  {journal} {Phys. Rev. D}\ }\textbf {\bibinfo {volume} {79}},\ \bibinfo
  {pages} {113010} (\bibinfo {year} {2009})},\ \Eprint
  {http://arxiv.org/abs/0904.3640} {arXiv:0904.3640 [hep-ph]} \BibitemShut
  {NoStop}%
\bibitem [{\citenamefont {Baldini}\ \emph {et~al.}(2018)\citenamefont {Baldini}
  \emph {et~al.}}]{Baldini:2018nnn}%
  \BibitemOpen
  \bibfield  {author} {\bibinfo {author} {\bibfnamefont {A.~M.}\ \bibnamefont
  {Baldini}} \emph {et~al.} (\bibinfo {collaboration} {MEG II}),\ }\href
  {\doibase 10.1140/epjc/s10052-018-5845-6} {\bibfield  {journal} {\bibinfo
  {journal} {Eur. Phys. J.}\ }\textbf {\bibinfo {volume} {C78}},\ \bibinfo
  {pages} {380} (\bibinfo {year} {2018})},\ \Eprint
  {http://arxiv.org/abs/1801.04688} {arXiv:1801.04688 [physics.ins-det]}
  \BibitemShut {NoStop}%
\bibitem [{\citenamefont {Chao}(2008)}]{Chao:2008iw}%
  \BibitemOpen
  \bibfield  {author} {\bibinfo {author} {\bibfnamefont {W.}~\bibnamefont
  {Chao}},\ }\href@noop {} {\  (\bibinfo {year} {2008})},\ \Eprint
  {http://arxiv.org/abs/0806.0889} {arXiv:0806.0889 [hep-ph]} \BibitemShut
  {NoStop}%
\bibitem [{\citenamefont {Chen}\ and\ \citenamefont
  {Nomura}(2019)}]{Chen:2019nud}%
  \BibitemOpen
  \bibfield  {author} {\bibinfo {author} {\bibfnamefont {C.-H.}\ \bibnamefont
  {Chen}}\ and\ \bibinfo {author} {\bibfnamefont {T.}~\bibnamefont {Nomura}},\
  }\href {\doibase 10.1103/PhysRevD.100.015024} {\bibfield  {journal} {\bibinfo
   {journal} {Phys. Rev. D}\ }\textbf {\bibinfo {volume} {100}},\ \bibinfo
  {pages} {015024} (\bibinfo {year} {2019})},\ \Eprint
  {http://arxiv.org/abs/1903.03380} {arXiv:1903.03380 [hep-ph]} \BibitemShut
  {NoStop}%
\bibitem [{\citenamefont {Fileviez~Perez}\ and\ \citenamefont
  {Wise}(2009)}]{FileviezPerez:2009ud}%
  \BibitemOpen
  \bibfield  {author} {\bibinfo {author} {\bibfnamefont {P.}~\bibnamefont
  {Fileviez~Perez}}\ and\ \bibinfo {author} {\bibfnamefont {M.~B.}\
  \bibnamefont {Wise}},\ }\href {\doibase 10.1103/PhysRevD.80.053006}
  {\bibfield  {journal} {\bibinfo  {journal} {Phys. Rev.}\ }\textbf {\bibinfo
  {volume} {D80}},\ \bibinfo {pages} {053006} (\bibinfo {year} {2009})},\
  \Eprint {http://arxiv.org/abs/0906.2950} {arXiv:0906.2950 [hep-ph]}
  \BibitemShut {NoStop}%
\bibitem [{\citenamefont {Bauer}\ and\ \citenamefont
  {Plehn}(2019)}]{Plehn:2017fdg}%
  \BibitemOpen
  \bibfield  {author} {\bibinfo {author} {\bibfnamefont {M.}~\bibnamefont
  {Bauer}}\ and\ \bibinfo {author} {\bibfnamefont {T.}~\bibnamefont {Plehn}},\
  }\href {\doibase 10.1007/978-3-030-16234-4} {\emph {\bibinfo {title} {{Yet
  Another Introduction to Dark Matter}: {The Particle Physics Approach}}}},\
  \bibinfo {series} {Lecture Notes in Physics}, Vol.\ \bibinfo {volume} {959}\
  (\bibinfo  {publisher} {Springer},\ \bibinfo {year} {2019})\ \Eprint
  {http://arxiv.org/abs/1705.01987} {arXiv:1705.01987 [hep-ph]} \BibitemShut
  {NoStop}%
\bibitem [{\citenamefont {No}\ \emph {et~al.}(2020)\citenamefont {No},
  \citenamefont {Tunney},\ and\ \citenamefont {Zaldivar}}]{No:2019gvl}%
  \BibitemOpen
  \bibfield  {author} {\bibinfo {author} {\bibfnamefont {J.~M.}\ \bibnamefont
  {No}}, \bibinfo {author} {\bibfnamefont {P.}~\bibnamefont {Tunney}}, \ and\
  \bibinfo {author} {\bibfnamefont {B.}~\bibnamefont {Zaldivar}},\ }\href
  {\doibase 10.1007/JHEP03(2020)022} {\bibfield  {journal} {\bibinfo  {journal}
  {JHEP}\ }\textbf {\bibinfo {volume} {03}},\ \bibinfo {pages} {022} (\bibinfo
  {year} {2020})},\ \Eprint {http://arxiv.org/abs/1908.11387} {arXiv:1908.11387
  [hep-ph]} \BibitemShut {NoStop}%
\end{thebibliography}%
\end{document}